\documentclass{article}
\usepackage[utf8]{inputenc}
\usepackage[super]{nth}
\usepackage{subfig}
\usepackage{graphicx}
\usepackage{placeins}
\usepackage[margin=1in]{geometry}
\usepackage{amsmath}
\usepackage{authblk}
\usepackage{appendix}
\usepackage{xcolor}
\usepackage{enumitem}

\usepackage[backend=bibtex,sorting=none,style=numeric-comp]{biblatex}
\bibliography{main}

\definecolor{Orange}{named}{orange}

\begin{document}

\title{A Grand Scan of the pMSSM Parameter Space for Snowmass 2021}

\author[1]{Jennet Dickinson}
\author[2]{Samuel Bein}
\author[3]{Sven Heinemeyer}
\author[4]{Joshua Hiltbrand}
\author[1]{Jim Hirschauer}
\author[5]{Walter Hopkins}
\author[6]{Elliot Lipeles}
\author[2]{Malte Mrowietz}
\author[4]{Nadja Strobbe}

\affil[1]{Fermi National Accelerator Laboratory, Batavia, IL 60563, USA}
\affil[2]{Universität Hamburg, 20148 Hamburg, Germany}
\affil[3]{Instituto de F\'isica Te\'orica (UAM/CSIC), 
Universidad Aut\'onoma de Madrid\\ 
Cantoblanco, 28049, Madrid, Spain}
\affil[4]{University of Minnesota, Minneapolis, MN 55455, USA}
\affil[5]{Argonne National Laboratory, Lemont, IL 60439, USA}
\affil[6]{University of Pennsylvania, Philadelphia, PA 19104, USA}

\date{July 12, 2022}

\maketitle

\section{Introduction}

It is well known that the Standard Model of particle physics (SM) cannot be the ultimate theory of fundamental particles and interactions. Most obviously, the SM does not incorporate gravity and explains neither the identity of the astronomically observed dark matter (DM) nor the observed multiplicities and hierarchies of interactions, flavor, and fermion generations. Of the many models proposed to
address the shortcomings of the SM, supersymmetry (SUSY) garners significant interest
because it simultaneously explains the finite mass of the recently discovered Higgs particle, provides a DM candidate, and allows more precise unification of the forces.

The minimal supersymmetric standard model (MSSM) \cite{FAYET1976159, FAYET1977489, FARRAR1978575, FAYET1979416, DIMOPOULOS1981150} has 105 parameters (in addition to the SM parameters) describing particle masses and interactions. In order to facilitate interpretation of experimental results within the MSSM framework, 
there are two main possibilities to deal with this large number of free parameters. One consists in assuming unification at some high scale, at which a certain soft SUSY-breaking mechanism is adopted. This can reduce the number of free parameters to as few as five, depending on the mechanism responsible for the values of the new particles' masses, referred to as the soft SUSY breaking mechanism. The most prominent example is the ``Constrained MSSM'' (cMSSM)~\cite{Chamseddine:1982jx,Barbieri:1982eh,Ibanez:1982ee,Hall:1983iz}. Another assumes the production of one set of SUSY particles at the electroweak (EW) scale with a fixed decay chain. This approach is called ``simplified model spectra'' (SMS) approach~\cite{sms2009,sms2012,cms2013}.
%these 120 free parameters have traditionally been reduced to five, by assuming relationships between MSSM parameters based on a choice of the SUSY breaking mechanism at high energy scale, in the form of the constrained MSSM (cMSSM) \cite{Chamseddine:1982jx,Barbieri:1982eh,Ibanez:1982ee,Hall:1983iz} or two parameters, by assuming pair production of a single SUSY particle (sparticle) with fixed decay chain, in the form of simplified model spectra (SMS) \cite{sms2009,sms2012,cms2013}. 
While these frameworks allow for efficient interpretation of results, they do so at the expense of sampling only a very small part of the phase space of the MSSM and potentially focusing on signatures that may not be realized in nature.

In recent years, the ATLAS and CMS collaborations~\cite{ATLAS_pMSSM,CMS_pMSSM} and theorists have attempted to ameliorate the limitations of interpretations based on the cMSSM and SMS by using the phenomenological MSSM (pMSSM)\cite{atlaspmssm2015,cmspmssm2016,pmssm2009,pmssm2012,Ambrogi_2018,AbdusSalam:2011fc}. The pMSSM reduces the full MSSM space to nineteen free parameters, specified at the electroweak (EW) scale, based on assumptions consistent with current experimental constraints rather than details of the soft SUSY breaking mechanism. 
%, including those from flavor, CP violation, and EW symmetry breaking.  
The parameters of the pMSSM and their definitions are listed in Table \ref{tab:parameters}.

\begin{table}[!htp]
\begin{center}
\begin{tabular}{|l|l|l|l|}
\hline
\textbf{Param.} & \textbf{Definition} & \textbf{Range} & \textbf{Sampling}\\
\hline\hline
$M_A$ & mass of pseudoscalar Higgs boson & 100 GeV - 25 TeV & Log\\
$\tan \beta$ & ratio of Higgs vevs & 1 - 60 & Log\\
$|\mu|$ & Higgs-higgsino mass parameter & 80 GeV - 25 TeV & Log\\
$|M_1|$ & bino mass parameter & 1 GeV - 25 TeV & Log\\
$|M_2|$ & wino mass parameter & 70 GeV - 25 TeV & Log\\
$M_3$ & gluino mass parameter & 200 GeV - 50 TeV & Linear \\
$m_{\tilde{L}^{1,2}}$ & \nth{1}, \nth{2} gen. left-handed slepton mass & 90 GeV - 25 TeV & Log\\
$m_{\tilde{R}^{1,2}}$ & \nth{1}, \nth{2} gen. right-handed slepton mass & 90 GeV - 25 TeV & Log\\
$m_{\tilde{L}^3}$ & \nth{3} gen. left-handed slepton mass & 90 GeV - 25 TeV & Log\\
$m_{\tilde{R}^3}$ & \nth{3} gen. right-handed slepton mass & 90 GeV - 25 TeV & Log\\
$m_{\tilde{q}^{1,2}}$ & \nth{1}, \nth{2} gen. left-handed squark mass & 200 GeV - 50 TeV & Linear\\
$m_{\tilde{u}^{1,2}}$ & \nth{1}, \nth{2} gen. right-handed $u$-type squark mass & 200 GeV - 50 TeV & Linear\\
$m_{\tilde{d}^{1,2}}$ & \nth{1}, \nth{2} gen. right-handed $d$-type squark mass & 200 GeV - 50 TeV & Linear\\
$m_{\tilde{q}^3}$ & \nth{3} gen. left-handed squark mass & 100 GeV - 50 TeV & Linear\\
$m_{\tilde{u}^3}$ & right-handed stop quark mass parameter & 100 GeV - 50 TeV & Linear\\
$m_{\tilde{d}^3}$ & right-handed sbottom quark mass parameter & 100 GeV - 50 TeV & Linear\\
$|A_\tau|$ & $\tau$ trilinear coupling & 1 GeV - 7 TeV & Log\\
$|A_b|$ & bottom trilinear coupling & 1 GeV - 7 TeV & Log\\
$|A_t|$ & top trilinear coupling & 1 GeV - $3 ( m_{\tilde{q}^3} m_{\tilde{u}^3} )^{1/2}$ & Log\\
\hline
\end{tabular}
\caption{The 19 parameters of the pMSSM, their allowed ranges, and sampling method in the Snowmass 2021 scan. Sampling methods are described in Sec.~\ref{sec:stepping}. All complex parameters are taken to be real. }
\label{tab:parameters}
\end{center}
\end{table}

Building on the methods of the CMS and ATLAS collaborations, we have developed a flexible framework for interpretation of SUSY sensitivity studies for future colliders in the framework of the pMSSM. We perform a \textit{grand scan} of the pMSSM parameter space that covers the logical OR of accessible ranges of many collider scenarios, including electron, muon, and hadron colliders at a variety of center of mass energies. This enables comparisons of sensitivity and complementarity of different future experiments, including both colliders and precision measurements in the Cosmological and Rare Frontiers. 

The scan procedure is described in detail in later sections, but we summarize the general approach here. The SM parameters required for spectrum generation are fixed to their experimentally determined values, except for the top quark mass, bottom quark mass, and strong coupling constant $\alpha_S$, which are sampled from Gaussian distributions based on their measured value and uncertainty as shown in Table \ref{tab:sm}. The ranges of values for pMSSM parameters are listed in Table \ref{tab:parameters}. The lower bounds on each parameter are chosen based on experimental constraints and phenomenology considerations. The upper bounds are chosen to be within reach of the sensitivity of a 100 TeV proton-proton collider.   

\begin{table}[!htp]
\begin{center}
\begin{tabular}{|l|l|l|l|}
\hline
\textbf{SM Param.} & \textbf{Definition} & \textbf{Measurement}  \\
\hline\hline
% https://pdg.lbl.gov/2019/tables/rpp2019-sum-quarks.pdf  
$m_t$ & top quark mass & $173.1 \pm 0.9$ GeV \\
$m_b$ & bottom quark mass & $4.18 _{-0.02}^{+0.03}$ GeV \\
$\alpha_s$ & strong coupling constant & $0.1181 \pm 0.0011$\\
\hline
\end{tabular}
\caption{SM parameters that are sampled in the pMSSM scan.  All other SM parameters are fixed to their experimentally determined values.}
\label{tab:sm}
\end{center}
\end{table}

The output of the scan for each selected pMSSM point includes the pMSSM parameter values, the values of SM parameters drawn from distributions in Table~\ref{tab:sm}, and the values of interesting observables computed for the  selected pMSSM point including the Higgs boson mass and couplings, the anomalous magnetic moment of the muon, and the dark matter composition and relic density.  The SUSY particle spectrum is obtained from the SPheno 4.0.5 \cite{Porod2003,Porod2012} spectrum generator, which also produces an SLHA file~\cite{slha2004} for generating signal events for each point. 
%Event generation and detector simulation is not discussed in this whitepaper. 
Section \ref{sec:mcmc} describes how the scan is performed by using a Markov chain Monte Carlo procedure to sample points from the 19-dimensional pMSSM parameter space. Section \ref{sec:postprocess} describes how properties of theoretical and experimental interest are calculated for each point. Section \ref{sec:results} reviews the scan coverage in terms of a selection of interesting physics processes and estimates the impact of future precision measurements on the pMSSM parameter space. Conclusions and an outlook are given in Section \ref{sec:conclusion}. 
%Appendix \ref{sec:coverage} 

\section{Markov chain Monte Carlo sampling procedure}
\label{sec:mcmc}

Because of the high dimensionality of the pMSSM and the large range of parameter values considered in the scan (Table \ref{tab:parameters}), the parameter space covered by this scan is extremely large. A Markov chain Monte Carlo (McMC) algorithm \cite{ohagan,markov,metropolis,hastings,berg} is used to explore the space in an efficient way, guided by a likelihood constructed from existing experimental results as described in Sec.~\ref{sec:likelihood}.

To begin the McMC scan, an initial pMSSM point is selected at random within the ranges specified in Table \ref{tab:parameters}. The SPheno spectrum generator is used to calculate the corresponding particle masses and decays. The Higgs sector is then replaced with that calculated by FeynHiggs 2.18.0 \cite{bahl2019precision,Bahl_2018,Bahl_2016,Hahn_2014,Frank_2007,Degrassi_2003,Heinemeyer_1999,Heinemeyer_2000}. If the resulting point is not a \textit{valid} pMSSM model, e.g., if there is a tachyon present in the spectrum, a new point is selected randomly and the process is repeated until a viable initial point is obtained. The McMC likelihood for the initial point is calculated as described in Section \ref{sec:likelihood}, and then the following steps are repeated until the desired number of points is selected:
\begin{enumerate}
    \item A new pMSSM point $\vec{x}'$ is obtained from the most recently accepted point ($\vec{x}$) by sampling randomly from the probability distribution given by the stepping functions $f_i(x_i)$, where $i$ is the parameter index. The form of the stepping function is discussed in detail in Section \ref{sec:stepping}.
    \item If any parameter value of $\vec{x}'$ does not fall within the ranges listed in Table \ref{tab:parameters}, return to step 1.
    \item The particle spectrum and Higgs sector quantities for $\vec{x}'$ are calculated with SPheno and FeynHiggs, respectively. If $\vec{x}'$ is not a valid model or the uncertainty on the calculated Higgs boson mass is larger than 5 GeV, return to step 1.
    \item Perform additional calculations required to compute the McMC likelihood as described in detail in Sec.~\ref{sec:likelihood}.
    \item The point is accepted with a probability of its likelihood ratio $L(\vec{x}')/L(\vec{x})>1$ and rejected otherwise, which allows the McMC to move away from potential local minima.
    %If the likelihood ratio $L(\vec{x}')/L(\vec{x})>1$, the point is accepted.  If $L(\vec{x}')/L(\vec{x})\leq1$, the point is rejected if the likelihood ration does not exceed a random number in the interval [0,1] and accepted otherwise, which allows the McMC to move away from potentially local minima.  
    \item If the point is rejected, return to step 1.
    \item If the point is accepted, save the point and repeat from step 1 taking $\vec{x}'$ as the new most recently accepted point.
\end{enumerate}

The final pMSSM scan comprises the union of 400 independent McMC scan threads, each starting from a unique random initial point. 

\subsection{Construction of the McMC likelihood}
\label{sec:likelihood}
In order to steer the McMC out of regions of parameter space that are  excluded by existing measurements, a likelihood function is constructed to quantify the compatibility of each generated pMSSM point with current measurements.  For each of the following observables, a contribution to the total likelihood is computed as described and included in the total likelihood:

\begin{enumerate}[label=(\alph*)]
    \item branching ratios for $B\rightarrow\tau\nu$, $D_s\rightarrow\tau\nu$, and $D_s\rightarrow\mu\nu$ as computed with SPheno 4.0.5;\label{itm:a}
    \item isospin symmetry breaking in the decay $\Delta_0(B^0\rightarrow K^{\star}\gamma$) and other $B$ meson branching ratios shown in Table~\ref{tab:superiso} as computed with Superiso 4.1~\cite{superiso2009};\label{itm:b}
    \item SM Higgs boson mass and couplings as computed with HiggsSignals 2.6.0~\cite{Bechtle_2014,Bechtle:2020uwn};\label{itm:c}
    \item limits on the production of heavy Higgs bosons decaying to a pair of $\tau$ leptons as computed with HiggsBounds 5.9.1 ~\cite{Bechtle_2010,bechtle2013recent,Bechtle_2014b,Bechtle_2015,Bechtle:2020pkv};\label{itm:d}
    \item and the anomalous magnetic moment of the muon as computed with GM2Calc~\cite{gm2calc2016,gm2calc2021} and used in half of the pMSSM scan (see below).\label{itm:e}
\end{enumerate}

For item \ref{itm:a}, the contribution of each observable to the total likelihood is a Gaussian likelihood function, with mean and width corresponding to the measured value and uncertainty, evaluated at the branching ratio value calculated by SPheno. 
\begin{table}[!htp]
\begin{center}
\begin{tabular}{|l|c|}
\hline
\textbf{Observable} & \textbf{N(meas.)}\\
\hline\hline
$\Delta_0(B^0\rightarrow K^\star\gamma)$ & 1\\
$BR(B^0\rightarrow K^{*0}\gamma)$ & 1\\
$BR(B_s\rightarrow \mu\mu)$ & 1\\
$BR(B_d\rightarrow \mu\mu)$ & 1\\
$BR(b\rightarrow s\gamma)$ & 1\\
$BR(b\rightarrow s\mu\mu)$ & 2 \\
$BR(b\rightarrow see)$ & 2 \\
\hline
\end{tabular}
\caption{$B$-physics observables included in the $\chi^2$ from Superiso. For the last two rows, measurements from two separate energy regions are included.}
\label{tab:superiso}
\end{center}
\end{table}
For item \ref{itm:b}, Superiso is used to calculate the values of nine additional $B$-physics observables, shown in Table~\ref{tab:superiso}, from the spectrum provided by SPheno. Superiso provides the compatibility of the calculated observables with existing measurements in the form of a $\chi^2$, which is incorporated into the likelihood using the number of measurements for each observable $n$ with the factor
\begin{equation}
    L = \frac{(\chi^2)^{(\frac{n}{2}-1)}}{2^\frac{n}{2}\Gamma(\frac{n}{2})}e^{-\chi^2/2}\text{.}
    \label{eq:chi2}
\end{equation}
For item \ref{itm:c}, HiggsSignals 2.6.0 returns a $\chi^2$ value, which is incorporated into the total likelihood according to Equation \ref{eq:chi2} with $n=107$, corresponding to the number of coupling and mass measurements used.  For item \ref{itm:d}, the likelihood corresponding to $H/A\rightarrow\tau\tau$ searches at the LHC is calculated by HiggsBounds 5.9.1 and included in the total McMC likelihood; selection based on exclusions from other heavy Higgs search channels is applied in a post processing step described in Sec.~\ref{sec:postprocess}. For item \ref{itm:e}, the difference between the standard model value and the pMSSM value for the anomalous muon magnetic moment ($\Delta a_\mu$) is calculated for each point by the GM2Calc package. In order to allow for sufficient statistics near both the SM value of $\Delta a_\mu=0$ and the measured central value of $\Delta a_\mu = 251 \times 10^{-11}$ \cite{gminus22021}, $\Delta a_\mu$ is included in the likelihood for only half of the scan threads. When included in the likelihood, the $\Delta a_\mu$ contribution is taken to be Gaussian with mean corresponding to the measured central value, with the same value ($251 \times 10^{-11}$) used for the width. This approach ensures that pMSSM regions near the measured value and the SM prediction are both populated by the scan. 

The McMC likelihood does not include the following observables, though their values are computed for each point to allow understanding of how the allowed pMSSM space relates to physical observables of interest:
\begin{itemize}
    \item Character and composition of the lightest neutralino, which is taken as the DM candidate;
    \item DM-nucleon cross sections;
    \item DM relic density $\Omega h^2$.
\end{itemize}

\subsection{Stepping in the McMC}
\label{sec:stepping}

Given pMSSM point $\vec{x}$, the stepping functions $f_i(x_i)$ are the probability distributions from which the next pMSSM point is chosen, where $i$ is the parameter index.  Two forms are considered for the stepping functions in the pMSSM scan. The first is denoted \textbf{linear}:
\begin{equation}
    f(x_i) = \text{Gaus} (\mu=x_i,\sigma=\sigma_0\times w) \; \text{.}
\end{equation}
Here, $w$ is the width of the parameter range allowed for parameter $x_i$ (see Table \ref{tab:parameters}), and $\sigma_0$ is a tunable step size parameter. 
The second stepping function is denoted \textbf{logarithmic}:
\begin{equation}
    f(x_i) = \exp \left[ \text{Gaus} (\mu=\ln|x_i|,\sigma=\sigma_0\times w)\right] \; \text{,}
\end{equation}
where $w$ and $\sigma_0$ have the same meanings as above.  In general,
the linear function provides approximately uniform coverage over the range of parameter values, while the logarithmic function ensures that lower parameter values are explored with finer granularity than higher ones, which is advantageous for certain parameters in the scan. First, the highest parameter values are inaccessible to many collider scenarios of interest for Snowmass 2021, and the use of log stepping in some or all parameters ensures that the scan is not overwhelmed by points only accessible at a 100 TeV proton-proton collider.  Second, with a higher concentration of lower sparticle masses, it becomes more likely to obtain so called ``compressed spectra'' for which small differences between masses of SUSY particles in the decay chain result in a more diverse array of experimental signatures including low missing transverse momentum, leptons with low transverse momentum, and long-lived particles.

Because the logarithmic stepping function is positive definite, each new point selected from this distribution will have positive sign. However, some pMSSM parameters can have negative values ($M_1$, $M_2$, $\mu$, $A_t$, $A_b$, $A_l$).  The sign of each parameter is therefore fixed to that of the randomly selected initial point, and the magnitude only is determined by the stepping function. The final combination of many scans with different initial points ensures that all sign combinations are explored.

A series of small-scale test scans (200k points) are generated to optimize the choice of stepping functions. In these tests, the logarithmic stepping function is taken as the default for all pMSSM parameters, and four values of the step size are tested: $\sigma_0 =$ 0.05, 0.10, 0.20, and 0.30. These scans are labelled by ``Log $\sigma_0$'' for the remainder of this section. An additional scan setup, ``Lin~0.05'', aims to sample a higher fraction of points where the strong SUSY sector is decoupled: here, the logarithmic stepping function ($\sigma_0=0.05$) is used for all but the squark and gluino mass parameters, which instead employ the linear stepping function with $\sigma_0=0.05$. The configurations of the stepping function test scans are listed in the top section of Table \ref{tab:width}. The middle section of this table reports the number of points sampled, accepted by the McMC, and accepted after post-processing (described in Section \ref{sec:postprocess}), as well as the total efficiency. The lower portion of Table \ref{tab:width} shows the fraction of points in two regions that are of special interest:
\begin{itemize}
    \item points with lightest squark or gluino mass exceeding 10 TeV represent a sample in which the strong SUSY sector is decoupled from electroweak production at the LHC and HL-LHC, and
    \item points with mass difference between the lightest SUSY particle (LSP) and the stop or gluino not exceeding $500$ GeV represent a sample of compressed spectra described above.
\end{itemize}
A high density of points in both regions is useful for Snowmass 2021 studies.

\begin{table}[!htp]
\begin{center}
\begin{tabular}{|r|r|r|r|r|r|}
\hline
& Log 0.05 & Log 0.10 & Log 0.20 & Log 0.30 & Lin 0.05 \\ 
\hline\hline
Stepping ($\tilde{q}$, $\tilde{g}$) & log & log & log & log & lin \\
Stepping (other) & log & log & log & log & log \\
Step width $\sigma_0$ & 5\% & 10\% & 20\% & 30\% & 5\% \\
\hline
Sampled points & 200,100 & 200,100 & 200,100 & 200,100 & 200,100 \\
McMC accepted points & 1,376 & 483 & 227 & 178 & 46,141 \\
Post-process accepted points & 345 & 124 & 53 & 46 & 23,064\\
Total Efficiency & 0.17\% & 0.06\% & 0.03\% & 0.02\% & 11.8\% \\ \hline
Strong decoupling & 34 & 22 & 18 & 13 & 3,917 \\
$\Delta m$(LSP – $\tilde{g}$)  $<$ 500 GeV & 4 & 2 & 0 & 0 & 13 \\
$\Delta m$(LSP – $\tilde{t}$)  $<$ 500 GeV & 2 & 0 & 0 & 0 & 9 \\
\hline
\end{tabular}
\caption{Results of the test scans with different stepping configurations.  As described in the text, the criteria for points with ``strong decoupling'' is that the squark and gluino masses exceed $10$ TeV.}
\label{tab:width}
\end{center}
\end{table}

\begin{figure}[htbp]
  \centering
  \includegraphics[width=0.6\textwidth]{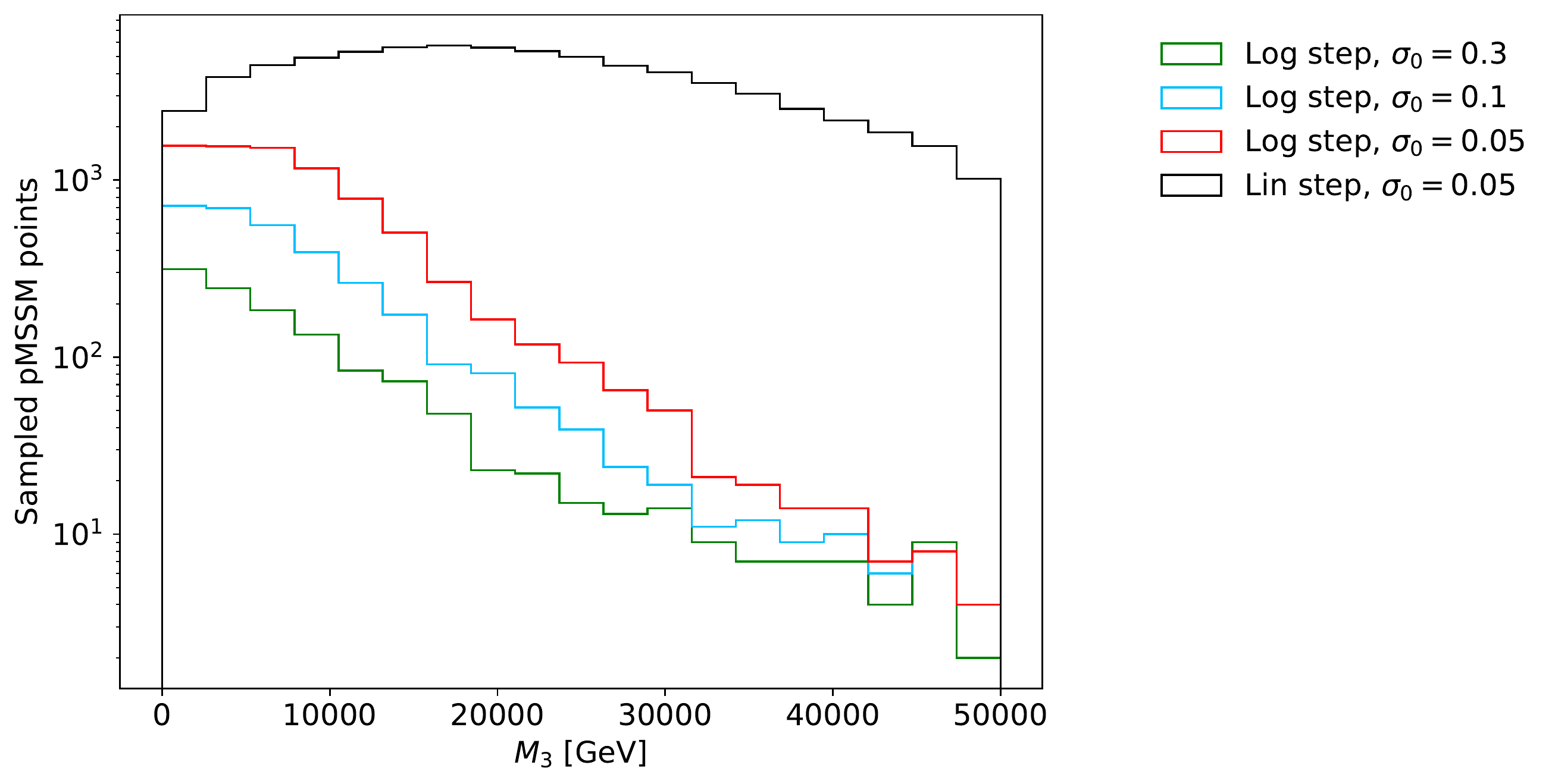} 
  \caption{Distribution of the gluino mass parameter $M_3$ for different stepping configurations.  The black distribution corresponds to Lin 0.05, which is chosen as the baseline. }
  \label{stepping}
\end{figure}

Figure \ref{stepping} shows the behavior of the different stepping configurations for an example parameter: $M_3$.  For the logarithmic stepping scenarios, an increase in $\sigma_0$ corresponds to an increase in the slope of the distribution, and therefore to the fraction of events with high squark masses. The difference in the integral of each distribution corresponds to the difference in acceptance efficiency of the McMC for the different configurations. 

The Lin 0.05 configuration results in the highest fraction of points with squark and gluino masses above 10 TeV, as seen in Table \ref{tab:width}. Largely for this reason, the Lin 0.05 test scan is deemed the most optimal and is used as the final stepping configuration. In order to maximize the statistics of the final scan, the results of the four non-optimal test scans presented in this section are combined with the results of the final configuration. However, the contribution from these test scans is small, and the overall scan behavior follows that of Lin 0.05. 

\FloatBarrier
\section{Selection of sampled points}
\label{sec:postprocess}

Following the sampling of pMSSM points according to the procedure described in the previous section, a post-processing step is applied to further remove experimentally disfavored regions of the parameter space from the scan. A scan point is excluded if it satisfies one or more of the following criteria:
\begin{enumerate}[label=(\alph*)]
    \item the LSP of the spectrum is not a neutralino,\label{itm:a}
    \item the point is excluded at 95\% CL by one ore more LHC Higgs searches as calculated by HiggsBounds,\label{itm:b}
    \item the point is excluded at 95\% CL by one or more LHC SUSY searches as calculated by SModelS 2.1\cite{Kraml:2013mwa,Ambrogi:2017neo,Ambrogi:2018ujg,Alguero:2020yhu,smodelsv2}\label{itm:c},
    \item the point is excluded at 90\% CL by dark matter measurements as calculated by MicrOMEGAs 5.2.7.a\label{itm:d}, \cite{micromegas2014,barducci2017collider,micromegas2009}.
\end{enumerate}

For item \ref{itm:b}, boolean constraints indicating whether a point is excluded by LHC Higgs searches are computed by HiggsBounds and applied at the post-processing step.  HiggsBounds tests each of the Higgs bosons individually, where in each case the potentially strongest bound is applied. A pMSSM point is rejected if it is excluded at 95\% CL by at least one of the applied Higgs boson search channels.  This is in addition to the inclusion of the HiggsBounds likelihood for the $H/A\rightarrow\tau\tau$ search directly in the McMC likelihood, as described in Sec.~\ref{sec:likelihood}.
%the four LHC Higgs searches with highest expected sensitivity to that point. 
For item \ref{itm:c}, SModelS computes exclusion limits from LHC SMS searches for each pMSSM point by decomposing each point into all relevant SMS models, computing the cross section of the simplified model processes, and then comparing the cross section of each simplified model process to the experimental limit. The pMSSM point is rejected if any simplified model process is excluded at 95\% CL. 
For item \ref{itm:d}, the MicrOMEGAs package calculates exclusion limits from $Z\rightarrow$ invisible,  DM direct detection experiments \cite{micromegas2009}, and searches for sparticle and DM production in $e^+e^-$ collisions from LEP \cite{barducci2017collider}. Points excluded by any of these measurements at 90\% CL are rejected. 

The DM relic density $\Omega h^2$ is also calculated at the post-processing step by MicrOMEGAs.  Though not included in the McMC likelihood, this observable can be used to further focus on regions of the pMSSM scan that are experimentally and theoretically motivated. 

An overview of the scanned pMSSM points is shown in Table \ref{tab:summary}. The first and second columns show the portions of the scan that do and do not use $\Delta a_\mu$ in the McMC likelihood, respectively. Note that the scan efficiency is much higher when $\Delta a_\mu$ is not included in the likelihood, but the fraction of points with $\Delta a_\mu$ near the measured value (last row) is small. The total scan statistics, corresponding to the sum of the first two columns, are shown in the final column of Table \ref{tab:summary}. 

\begin{table}[!htp]
\begin{center}
\begin{tabular}{|l|r|r|r|r|r}
\hline
& With $\Delta a_\mu$ & Without $\Delta a_\mu$ & Total \\ 
\hline\hline
Sampled points &  14,194,316 & 16,848,695 & 31,043,011\\
McMC accepted points & 156,599 & 6,800,642 & 6,957,241\\
Post-process accepted points & 44,444 & 3,312,480 & 3,356,924\\
Total Efficiency & 0.3\% & 20\% & 11\% \\ \hline 
Strong decoupling & 5,529 & 577,695 & 583,224\\
$\Delta m$(LSP, gluino)  $<$ 500 GeV & 24 & 2,412 & 2,436\\
$\Delta m$(LSP, stop)  $<$ 500 GeV & 14 & 1,480 & 1,494\\
$\Delta a_\mu$ within measured $\pm1\sigma$ & 19,540 & 3,212 & 22,752\\
\hline
\end{tabular}
\caption{Summary of the scanned pMSSM points. The criteria for points with strong decoupling is that the squark, gluino masses exceed 10 TeV. }
\label{tab:summary}
\end{center}
\end{table}

\FloatBarrier
\section{Results}
\label{sec:results}

Precision measurements of sensitive observables can be a powerful tool for constraining physics beyond the standard model.  This section presents studies of how measurements of Higgs boson properties, the anomalous magnetic moment of the muon, and dark matter at current and future experiments are expected to impact the allowed pMSSM parameter space. 

\subsection{Higgs boson properties}

\begin{figure}[htbp]
  \centering
  \subfloat{\includegraphics[width=0.95\textwidth]{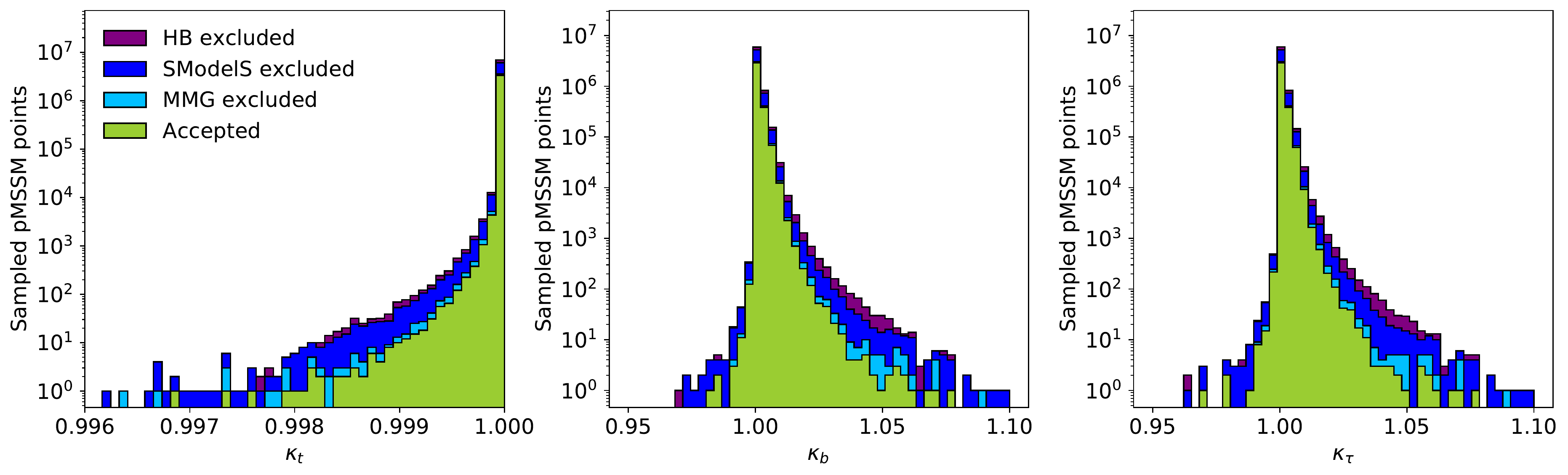}}
   \caption{Higgs couplings with top quark ($\kappa_t$), bottom quark ($\kappa_b$), and tau ($\kappa_\tau$) calculated by FeynHiggs for each pMSSM point.  Points accepted by the McMC but excluded by HiggsBounds (LHC Higgs searches) are shown in purple.  Of the remaining points, those excluded by SModelS (LHC SUSY searches) are shown in dark blue. And of the remaining points, those excluded  by MicrOMEGAs (DM measurements) are shown in light blue. The green distribution shows the points that are fully accepted after all post-processing steps.}
  \label{hquark}
\end{figure}

Higgs boson couplings for each pMSSM point are calculated by FeynHiggs and studied in the $\kappa$-framework \cite{kappa,kappa2}, where for particle $p$ with Higgs coupling $\lambda_p$
\begin{align}
    \kappa_p = \frac{\lambda_p}{\lambda_{p,\text{SM}}}~.
\end{align}
The Higgs boson coupling to quarks (top, bottom, and charm) and leptons (tau and muon) are found to peak sharply the expected SM value of $\kappa=1$. This is consistent with the LHC measurements to date, which enter the McMC likelihood through the contribution from HiggsSignals. The couplings to the top quark, bottom quark, and tau lepton are shown for each pMSSM point in Figure \ref{hquark}. The sampled pMSSM points that are excluded in the post-processing step described in Sec.~\ref{sec:postprocess} are shown separately from the final accepted sample. Points accepted by the McMC but excluded at 95\% CL by HiggsBounds (LHC Higgs searches) are shown in purple.  Of the remaining points, those excluded at 95\% CL by SModelS (LHC SUSY searches) are shown in dark blue. And of the remaining points, those excluded at 90\% CL by MicrOMEGAs (DM measurements) are shown in light blue. The green distribution shows the points that are fully accepted after all post-processing steps. The distributions of $\kappa_c$ and $\kappa_\mu$ (not shown) closely resemble those of $\kappa_t$ and $\kappa_\tau$, respectively. 

The distributions found for $\kappa_t$, $\kappa_b$, and $\kappa_\tau$ are consistent with deviations expected for tree-level contributions from the MSSM Higgs sector (a type-II two-Higgs-doublet model) rather than genuine SUSY loop contributions to the various Higgs couplings.  This observation is consistent with previous studies including Ref.~\cite{Bahl:2020kwe}.
The deviations from unity scale with $\sin(\beta - \alpha) + \cos(\beta - \alpha)/\tan\beta$ for $\kappa_t$ and with $\sin(\beta - \alpha) - \cos(\beta - \alpha) \tan\beta$ for $\kappa_{b,\tau}$ ($\alpha$ is the angle that diagonalizes the CP-even Higgs sector). For $M_A \to \infty$, one finds $\beta - \alpha \to \pi/2$ and $\sin(\beta - \alpha) \to 1$. Consequently, for $\tan\beta > 1$ (as in this scan) one finds smaller deviations from unity for $\kappa_t$ and larger opposite sign deviations for $\kappa_{b,\tau}$.
The pseudoscalar component of each Higgs-fermion coupling is also calculated, but is not found to deviate from zero in any of the sampled pMSSM points.
Taking into account the size of the deviations of $\kappa_{t,b,\tau}$ from unity and the expected experimental precision at current and future experiments in the various $\kappa$ determinations, $\kappa_b$ is expected to provide most sensitivity to deviations from the SM.

\begin{figure}[htbp]
  \centering
  \subfloat{\includegraphics[width=0.45\textwidth]{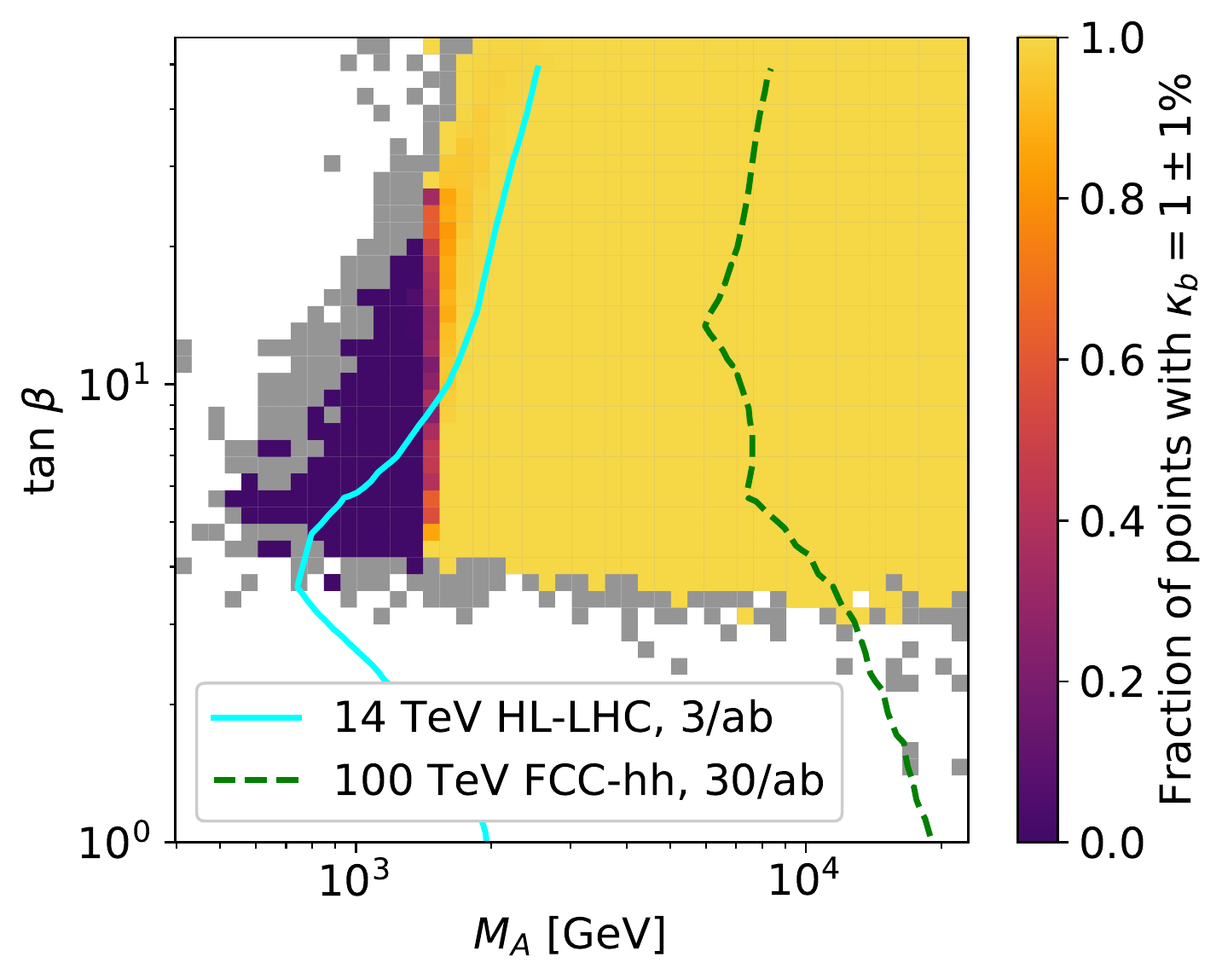}}
  \subfloat{\includegraphics[width=0.45\textwidth]{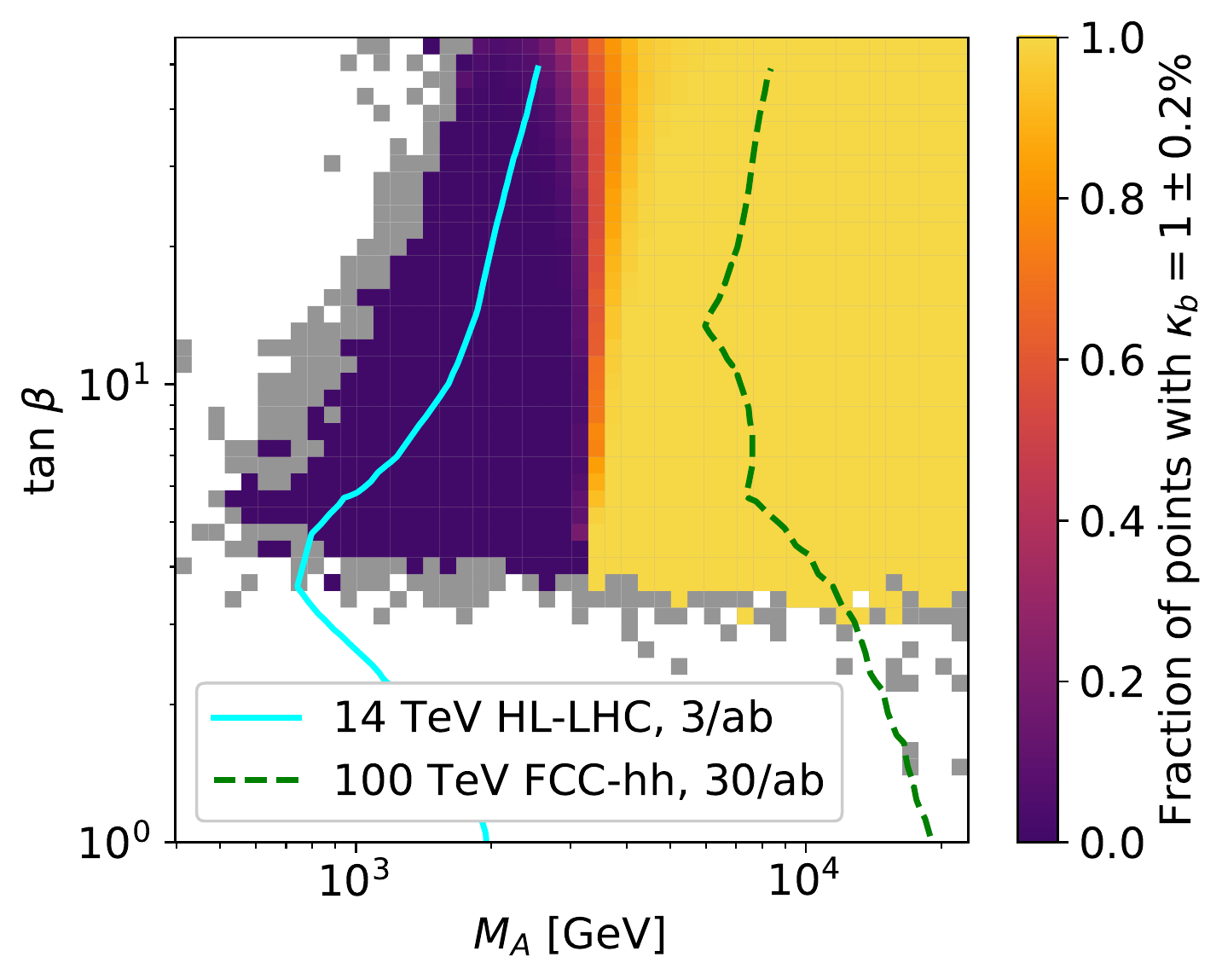}}
   \caption{Fraction sampled pMSSM points with $\kappa_b$ within $\pm 1\%$ (left) and $\pm 0.2\%$ (right) of the SM expectation, shown as a function of $\tan\beta$ and $M_A$. The $\pm 1\%$ range roughly corresponds to the 95\% CL precision expected for the combination of FCC-ee/eh/hh~\cite{de_Blas_2020}.  The projected limits from direct searches for heavy MSSM Higgs bosons at the HL-LHC (cyan) and the FCC-hh (green) are overlaid.  }
  \label{hbb-cuts-2d}
\end{figure}

Figure~\ref{hbb-cuts-2d} shows the fraction of the sampled pMSSM points in the $M_A$--$\tan\beta$ plane having $\kappa_b$ within $\pm 1\%$ of unity (on the left), which roughly corresponds to the 95\% CL precision expected for the combination of FCC-ee/eh/hh~\cite{de_Blas_2020}, and within $\pm 0.2\%$ of unity (on the right). The gray region is excluded by current searches for heavy MSSM Higgs bosons. The $1\%$ limit cuts away most points below $M_A \le 1500$~GeV, setting an indirect lower bound on the MSSM heavy Higgs mass scale. A stronger bound can be found for the increased precision of $0.2\%$, where $M_A$ less than $\sim 3$~TeV would only be allowed in a small fraction of models, all with large $\tan\beta$ values. The projected limits from direct searches for heavy MSSM Higgs bosons at the HL-LHC (cyan) and the FCC-hh (green) are overlaid.  Even with the full precision on $\kappa_b$ expected at the FCC-hh, the indirect limit is far less powerful than direct searches. 

\FloatBarrier
\subsection{Anomalous magnetic moment of the muon}

\begin{figure}[htbp]
  \centering
  \includegraphics[width=0.6\textwidth]{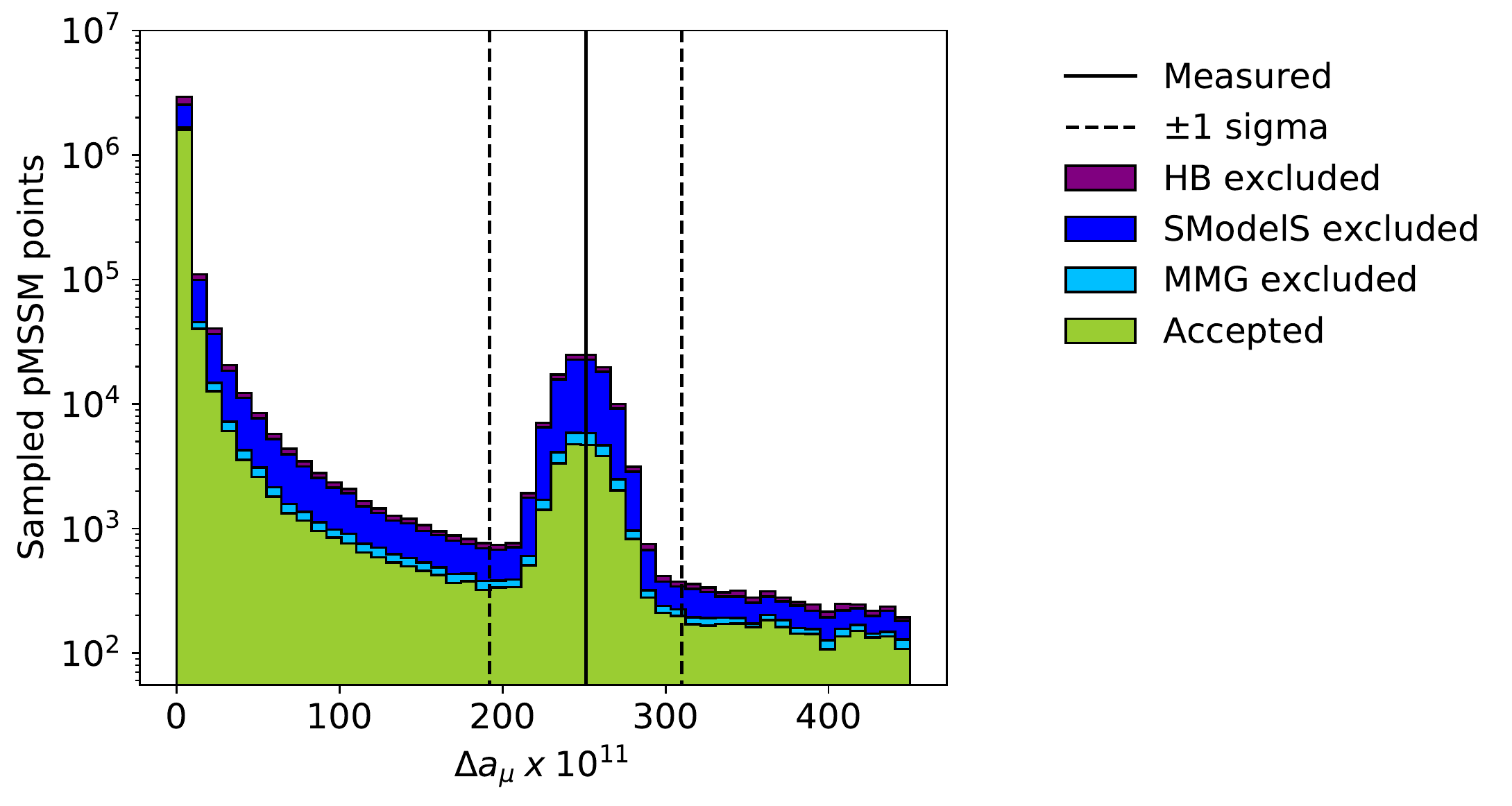}
   \caption{Distribution of $\Delta a_\mu$ for the sampled pMSSM points. The predicted Standard Model value corresponds to $\Delta a_\mu=0$, and the measured central value and uncertainties indicated by vertical lines.  Points accepted by the McMC but excluded by HiggsBounds (LHC Higgs searches) are shown in purple.  Of the remaining points, those excluded by SModelS (LHC SUSY searches) are shown in dark blue. And of the remaining points, those excluded by MicrOMEGAs (DM measurements) are shown in light blue. The green distribution shows the points that are fully accepted after all post-processing steps.}
  \label{delta_amu}
\end{figure}

Recent measurements of the anomalous muon magnetic moment $\Delta a_\mu$ \cite{gminus22021} that disagree with established SM predictions~\cite{Aoyama:2020ynm} at the level of 4.2 standard deviations have renewed interest in possible physics beyond the SM contributing to this observable. The McMC likelihood of the present pMSSM scan is constructed to populate the region near the measured value as well as the value predicted within the SM $(\Delta a_\mu=0)$. The distribution of $\Delta a_\mu$ for the sampled pMSSM points shown in Figure \ref{delta_amu} exhibits the desired two-peak structure. Though much of the peak at the measured value is excluded by SModelS, approximately 10\% of points passing all selection have $\Delta a_\mu$ within $1\sigma$ of the measurement (denoted by vertical lines).

\begin{figure}[htbp]
  \centering
  \subfloat{\includegraphics[width=0.45\textwidth]{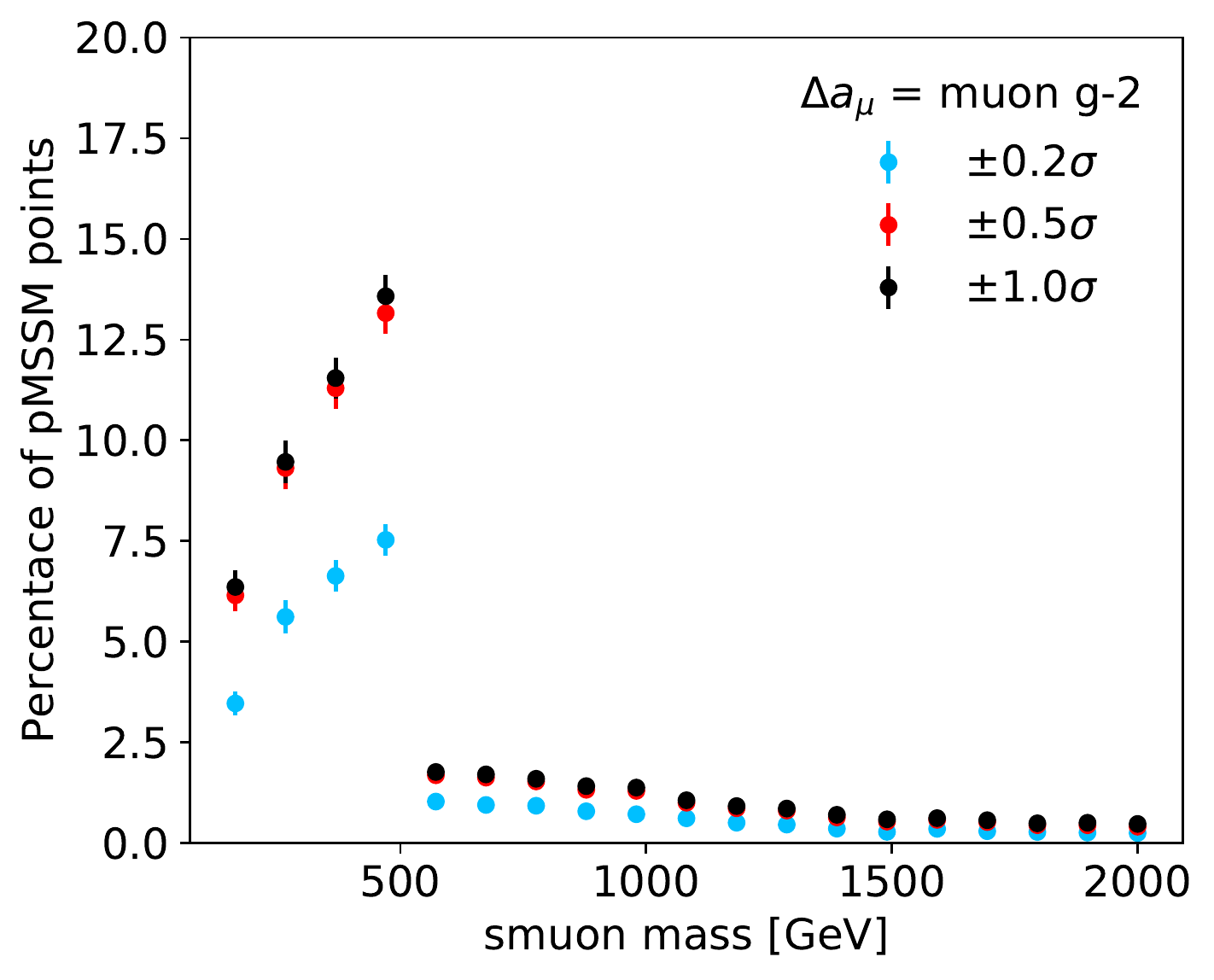}}
   \caption{Fraction of scanned pMSSM points with $\Delta a_\mu$ within $\pm1\sigma$ of the measured value, as a function of the smuon mass. The black line corresponds to the measured 2021 uncertainty $\sigma=59\times 10^{-11}$. The red and blue lines correspond to a reduction in the total uncertainty by a factor of two and five, respectively.}
  \label{delta_amu-cuts}
\end{figure}

In order to achieve a value of $\Delta a_\mu$ near the measured value of $\Delta a_\mu = 251 \times 10^{-11}$, SUSY contributions involving smuon-neutralino or muon sneutrino-chargino loops are required. The corresponding particle masses should not be too high, not to suppress these loop contributions, leading typically to a light smuon (see, e.g., \cite{Chakraborti:2021dli}).
The fraction of sampled pMSSM points with $\Delta a_\mu$ within $\pm 1 \sigma$ of the measured value is shown in Figure \ref{delta_amu-cuts} as a function of the smuon mass. Three values of $\sigma$ are considered: the uncertainty on the 2021 Muon $g-2$ measurement ($59\times 10^{-11}$), the projected uncertainty at the end of running of the Muon $g-2$ experiment (a factor of two improvement in over 2021 in the combined experimental and theoretical uncertainty), and a potential future precision benchmark of five times higher precision than the 2021 result. The fraction of sampled points satisfying the $\pm 1\sigma$ criterion is high for low values of the smuon mass ($<500$ GeV).

\FloatBarrier
\subsection{Electroweakino dark matter}

\begin{figure}[htbp]
  \centering
  \subfloat[Mass of neutralino vs. chargino]{\includegraphics[width=0.95\textwidth]{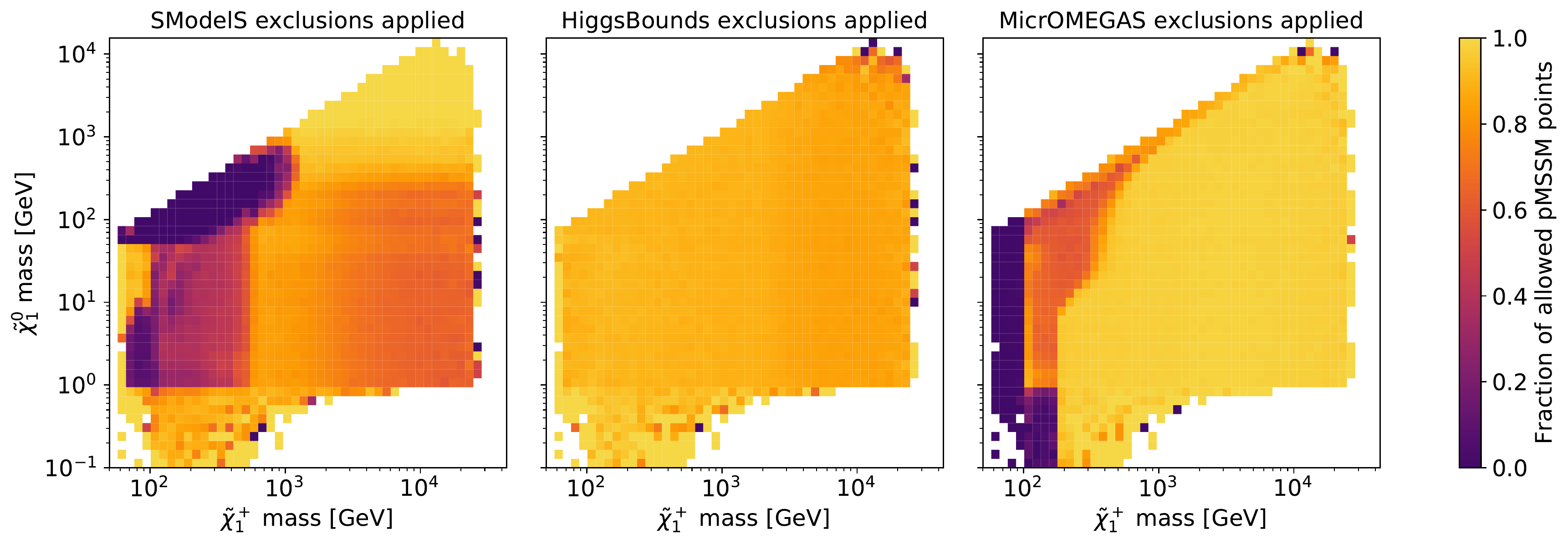}}
  \\
  \subfloat[DM relic density vs. neutralino mass]{\includegraphics[width=0.95\textwidth]{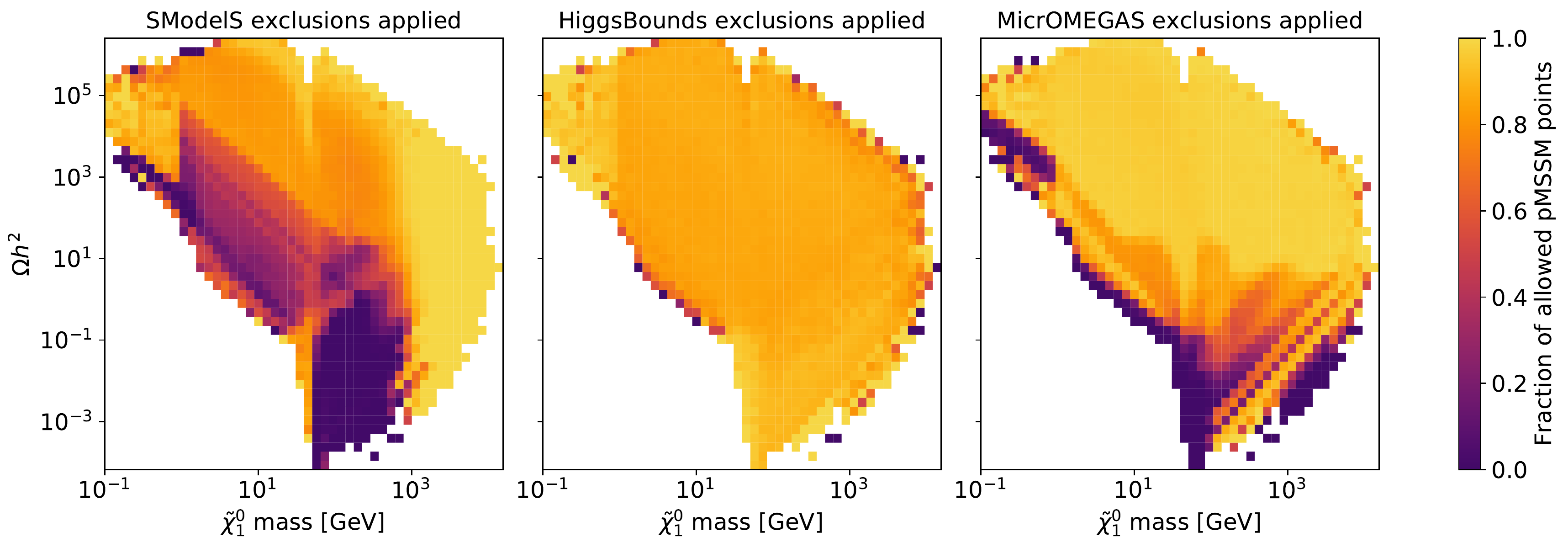}} \\
  \subfloat[Spin-independent neutralino-neutron cross section vs. neutralino mass]{\includegraphics[width=0.95\textwidth]{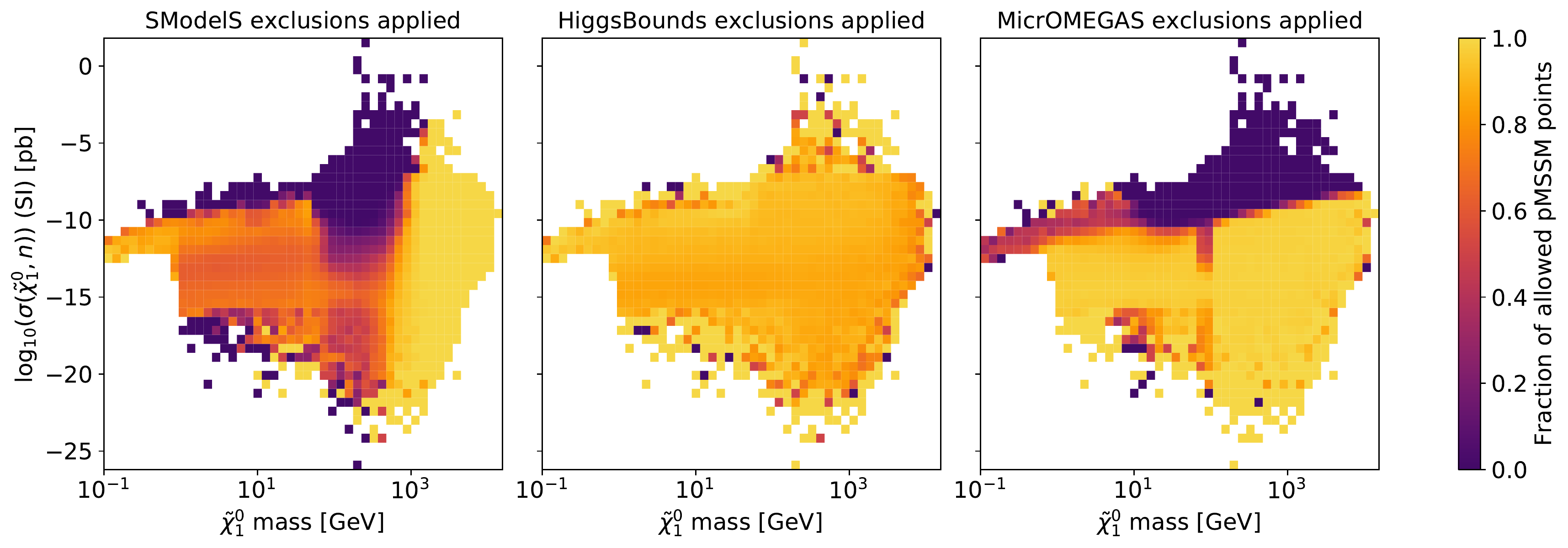}}
   \caption{Fraction of sampled pMSSM points after applying limits from existing measurements: from left to right, LHC SMS searches (SModelS), BSM Higgs searches (HiggsBounds), and DM/LEP measurements (MicrOMEGAs). Shown as a function of (a) neutralino vs. chargino mass, (b) relic density vs. neutralino mass, and (c) the spin-independent neutralino-neutron cross section vs. neutralino mass. The Planck measurement of $\Omega h^2$ is shown as a black line in (b). }
  \label{exclusions-2d}
\end{figure}

For each pMSSM point in the scan, the dark matter particle is taken to be the lightest neutralino in the model, $\tilde{\chi}^0_1$. The DM relic density ($\Omega h^2$) and DM-nucleon cross sections are calculated by MicrOMEGAs. Figure \ref{exclusions-2d} shows the fraction of pMSSM points surviving after limits from existing searches have been applied: from left to right, points are removed if they are excluded by LHC SMS searches (SModelS), BSM Higgs searches (HiggsBounds), and DM/LEP measurements (MicrOMEGAs). The three rows show the impact on different variable distributions: Figure \ref{exclusions-2d} (a) shows the lightest neutralino vs. lightest chargino mass plane, (b) shows the relic density vs. DM mass, and (c) shows the spin-independent DM-neutron cross section vs. DM mass. 

LHC searches for heavy stable charged particles (HSCPs) provide strong exclusions, but because HSCPs arise in the MSSM electroweakino sector primarily through small mass splittings between charged and neutral winos, these exclusions should be interpreted qualitatively rather than quantitatively.\footnote{The exact impact of LHC HSCP searches on the allowed wino-like dark matter (as well as the calculated DM mass) depends on the precise determination of the on-shell mass spectrum in the wino sector, which requires higher-order corrections that depend on the hierarchy of $M_2$, $M_1$ and $\mu$. These corrections so far cannot be taken into account precisely in automated way (see the discussion in \cite{Chakraborti:2021kkr}).} 

In the left panel of Figure \ref{exclusions-2d}a, HSCP searches at the LHC exclude nearly all sampled points in the region near $m(\tilde{\chi}^0_1)\sim m(\tilde{\chi}^+_1) \in 100-1000$ GeV. 
The excluded region at low neutralino mass near $m(\tilde{\chi}^+_1)\sim100$ GeV, where $<10\%$ of pMSSM points survive, corresponds to the exclusion from searches for chargino pair production with $\tilde{\chi}^\pm_1\rightarrow W^\pm\tilde{\chi}^0_1$.  The exclusion in the dark pink region from $m(\tilde{\chi}^0_1)\sim100-500$ GeV, which corresponds to 40-50\% of points remaining, is dominated by searches for the following final states:
\begin{itemize}
    \item $\tilde{\chi}^+_1\tilde{\chi}^-_1$ production with $\tilde{\chi}^\pm_1\rightarrow W^\pm\tilde{\chi}^0_1$;
    \item $\tilde{\chi}^\pm_1\tilde{\chi}^0_2$ production with $\tilde{\chi}^\pm_1\rightarrow W^\pm\tilde{\chi}^0_1$ and $\tilde{\chi}^0_2\rightarrow H\tilde{\chi}^0_1$;
    \item $\tilde{\chi}^\pm_1\tilde{\chi}^0_2$ production with $\tilde{\chi}^\pm_1\rightarrow W^\pm\tilde{\chi}^0_1$ and $\tilde{\chi}^0_2\rightarrow Z\tilde{\chi}^0_1$.
\end{itemize}
The right panel of Figure \ref{exclusions-2d}a shows the exclusions from MicrOMEGAs, which are dominated by mass limits from LEP (chargino mass $< 100$ GeV) and limits from $Z\rightarrow$ invisible (chargino mass $\sim 100-300$ GeV).

Figure \ref{exclusions-2d}b shows the impact of existing measurements as a function of DM relic density and mass. The Planck measurement of the DM relic density \cite{planck}, $\Omega h^2 = 0.120\pm0.001$, is indicated by a horizontal line in each panel.  The pMSSM scan populates values of this parameter over a very wide range compared to the experimental precision. The left panel shows the fraction of points surviving SModelS. The Z-funnel, where the Z boson contributes resonantly to the DM annihilation cross section, is clearly visible.  Points with $m(\tilde{\chi}^0_1) > m_Z$ and relic density less than $\sim 0.1$ (dark purple region) are excluded by HSCP searches. The excluded diagonal strip at $m(\chi^0_1)<100$ GeV corresponds to searches for slepton pair production.  The right panel shows that points with DM mass near $m_Z/2$ and relic density less than $\sim 0.1$ are excluded by MicrOMEGAS (specifically, DM direct detection experients and mass limits from LEP).  Points with DM mass less than $\sim1$ GeV are excluded by limits from $Z\rightarrow$ invisible for for lower relic densities.

Finally, Figure \ref{exclusions-2d}c shows exclusions in terms of the DM candidate mass and the spin-independent DM-neutron cross section (the DM-proton cross section yields a similar distribution).  Few points are sampled with DM mass below 1 GeV: this corresponds to the lower limit on $M_1$ used in the scan, and points with lower neutralino masses require some mixing between bino and other electroweakino states. In the left panel, LHC HSCP searches exclude points with high cross section, with the strongest limits in the range 100-1000 GeV. In the right panel, the exclusion due to DM direct detection experiments is clearly visible at high cross section values in the region above $m(\chi_1^0)\sim10$ GeV. 

\begin{figure}[htbp]
  \centering
  \includegraphics[width=0.9\textwidth]{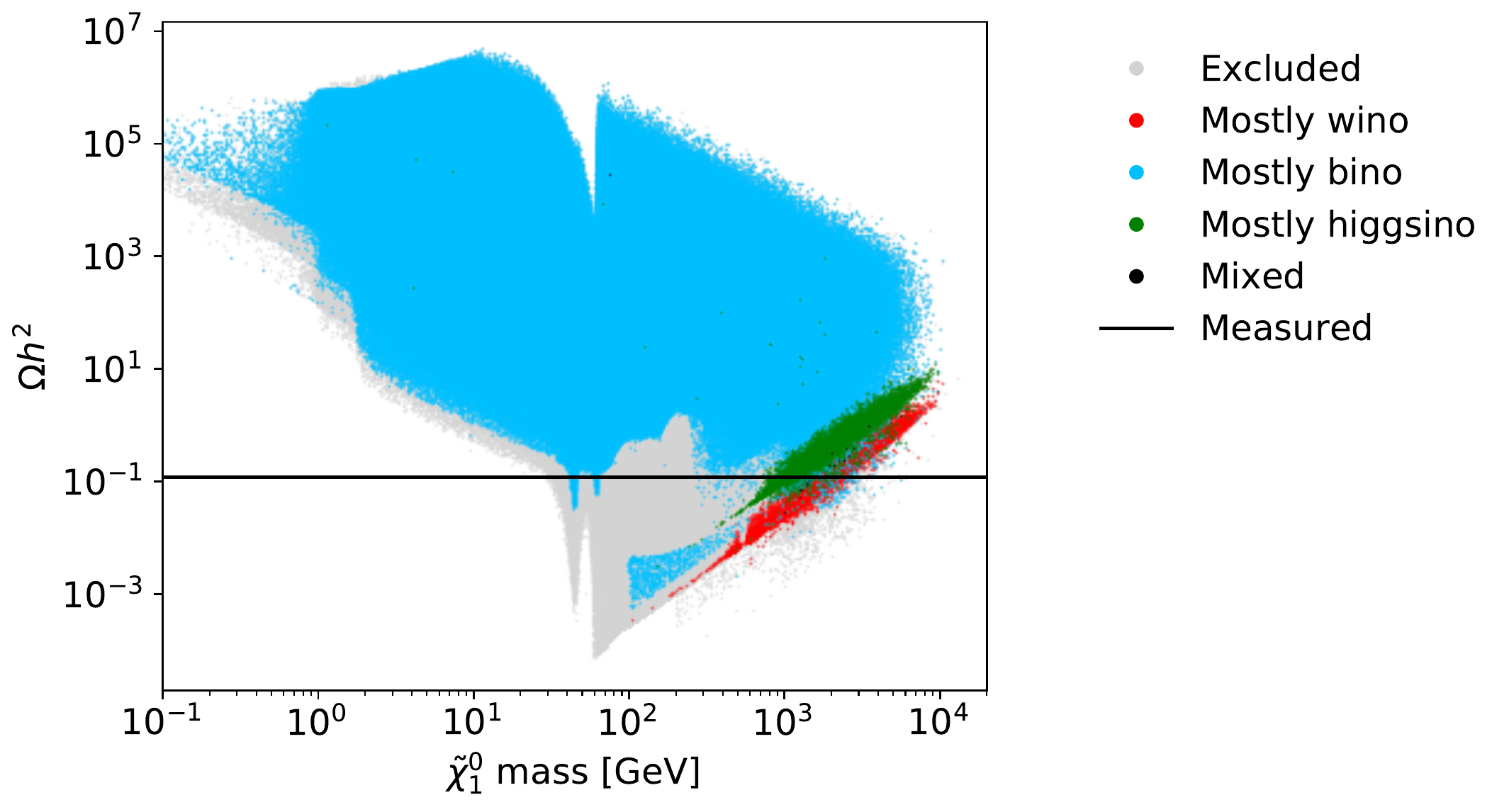}
   \caption{Distribution of sampled pMSSM points as function of the DM relic density vs. DM candidate mass. Points excluded by existing measurements (as implemented in SModelS, HiggsBounds, and MicrOMEGAS) are shown in light gray. The remaining points are broken down by the electroweakino composition of the DM candidate. The DM candidate is considered mostly pure if a single electroweakino comprises $>80\%$ of the admixture, and mixed otherwise.}
  \label{ewkinodm2d}
\end{figure}

The DM candidate particle is a superposition of electroweakino states (bino, wino, and higgsino). Each pMSSM point is labelled by the electroweakino comprising the largest component of $\tilde{\chi}^0_1$, thereby separating the scan into samples with bino-, wino-, and higgsino-like DM. Table \ref{tab:ewkino} shows the breakdown of pMSSM points by the composition of the DM candidate. The two-dimensional distribution of the DM candidate mass and relic density is shown for each component sample in Figure \ref{ewkinodm2d}. The sampled points with very high relic density are dominated by models with bino-like DM. Wino-like DM, on the other hand, is concentrated at low values of $\Omega h^2$. Near the measured value of the relic density, wino-, bino-, and higgsino-like DM all contribute.  Points that are excluded by SModelS, HiggsBounds, and MicrOMEGAS are shown in light gray. 
\begin{table}[!htp]
\begin{center}
\begin{tabular}{|l|r|r|}
\hline
& McMC accepted & Post-process accepted \\ 
\hline\hline
Mostly wino & 872,256 & 188,576 \\
Mostly bino & 5,513,680 & 3,071,103 \\
Mostly higgsino & 506,469 & 97,222 \\ \hline
Mixed wino/bino & 172 & 6 \\
Mixed bino/higgsino & 9,219 & 2 \\
Mixed wino/higgsino & 8,100 & 1 \\
Other & 47,445 & 14 \\
\hline
\end{tabular}
\caption{Number of scanned pMSSM points with different composition of the DM candidate. The DM candidate is considered mostly pure if a single electroweakino comprises $>80\%$ of the admixture and mixed if two components each contribute $> 40\%$.}
\label{tab:ewkino}
\end{center}
\end{table}

\begin{figure}[htbp]
  \centering
  \subfloat[]{\includegraphics[width=0.95\textwidth]{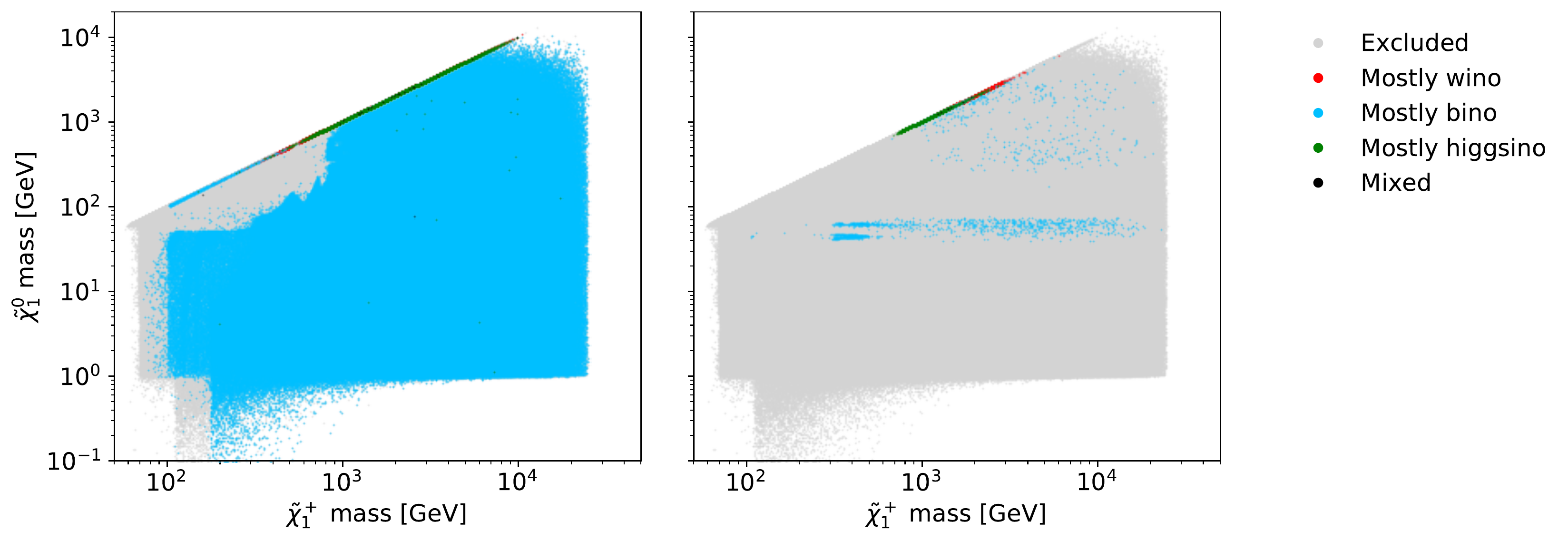}} \\
  \subfloat[]{\includegraphics[width=0.95\textwidth]{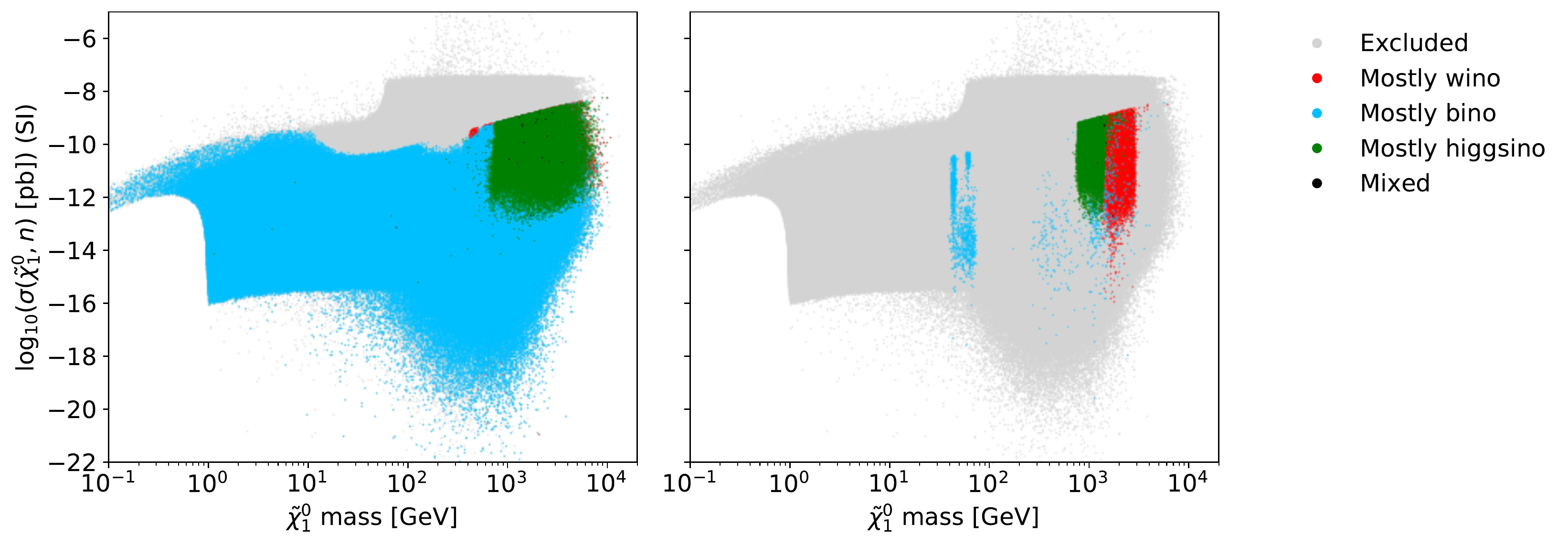}}
   \caption{Distribution of (a) the chargino and neutralino mass, and (b) the DM candidate mass and DM-neutron spin-independent cross section for the sampled pMSSM points. In the left panels, points excluded by existing measurements (as implemented in SModelS, HiggsBounds, and MicrOMEGAS) are shown in light gray. The remaining points are broken down by the electroweakino composition of the DM candidate. The right panels show the same distribution with the additional criterion that points with $\Omega h^2$ differing from the Planck measurement by more than 50\% are also colored gray.}
  \label{ewkinodm2d2}
\end{figure}

Figure \ref{ewkinodm2d2}a shows the distribution of neutralino vs. chargino mass, with each point color-coded according to the DM composition. In the left panel, points excluded by SModelS, HiggsBounds, and MicrOMEGAS are shown in light gray, as in Figure \ref{ewkinodm2d}. Because wino- and bino-like dark matter tend to have compressed spectra, where $m(\tilde{\chi}^0_1)\sim m(\tilde{\chi}^+_1)$, these points are clustered on the diagonal. The right panel shows the same distribution, but with the additional criterion that points with $\Omega h^2$ differing from the Planck measurement by more than 50\% are also colored gray. 

Figure \ref{ewkinodm2d2}b shows the distribution of the DM candidate mass and DM-neutron spin-independent cross section broken down by the DM composition. The colored points in the left panel correspond to models that are not excluded by existing measurements (SModelS, HiggsBounds, and MicrOMEGAS), while in the right panel the colored points are further required to be within 50\% of the measured relic density. 

The requirement that models yield a DM relic density within 50\% of the Planck measurement rules out all sampled points with DM mass less than $\sim 25$ GeV (dominantly bino-like). The surviving points include wino, bino, or higgsino dark matter candidates. 

\FloatBarrier

\section{Conclusion}
\label{sec:conclusion}

An extensive scan of the pMSSM parameter space has been completed for Snowmass 2021. The scanned parameter space covers ranges that will be accessible ranges to many future collider scenarios, including electron, muon, and hadron colliders at a center of mass energies up to 100 TeV.  The choice of a Markov chain Monte Carlo sampling method ensures that the 19 dimensional parameter space is explored efficiently. The inclusion of current measurements in the likelihood ensures that space that is experimentally disfavored is not prioritized. Additional selection at the post-processing stage accounts for the effects of LHC Higgs measurements, LHC searches for SUSY and other new phenomena, and DM measurements. The impact of future precision measurements on Higgs couplings, $\Delta a_\mu$, and DM quantities is presented. 

The next steps for this ongoing Snowmass pMSSM effort include performing studies with simulated events in order to quantitatively assess the sensitivity of selected future colliders in the context of the pMSSM. Studies will be based on common tools, such as Pythia8 for event generation and Delphes3 for detector simulation, and will employ implementations of recent searches for BSM physics at the LHC with optimization for future colliders.  Following these studies, pMSSM points not excluded by current or future experiments can be used to identify promising new regions in the space of experimental observables (e.g. particle momenta and invariant masses) in order to guide design of future analyses and detectors. We are exploring the use of machine-learning-based clustering techniques based on $k$-means~\cite{kmeans} and density-based clustering~\cite{dbscan} to identify categories and group pMSSM models with similar experimental signatures together. 

\FloatBarrier

\section*{Acknowledgements}
This research is supported in part by the DOE grants DE-SC0007901 and DE-SC0011845. 
This manuscript has been authored in part by Fermi Research Alliance, LLC under Contract No. DE-AC02-07CH11359 with the U.S. Department of Energy, Office of Science, Office of High Energy Physics.

\printbibliography
\clearpage

\appendix 
\section{Coverage of the scan}
\label{sec:coverage}

Throughout this section, sampled pMSSM points that are excluded at post-processing are shown separately from the final accepted sample. Points accepted by the McMC but excluded at 95\% CL by HiggsBounds (LHC Higgs searches) are shown in purple.  Of the remaining points, those excluded at 95\% CL by SModels (LHC SUSY searches) are shown in dark blue. And of the remaining points, those excluded at 90\% CL by MicrOMEGAs (dark matter measurements) are shown in light blue. The green distribution shows the points that are fully accepted after all post-processing steps. 

\subsection{pMSSM parameters}

\begin{figure}[htbp]
  \centering
  \includegraphics[width=0.7\textwidth]{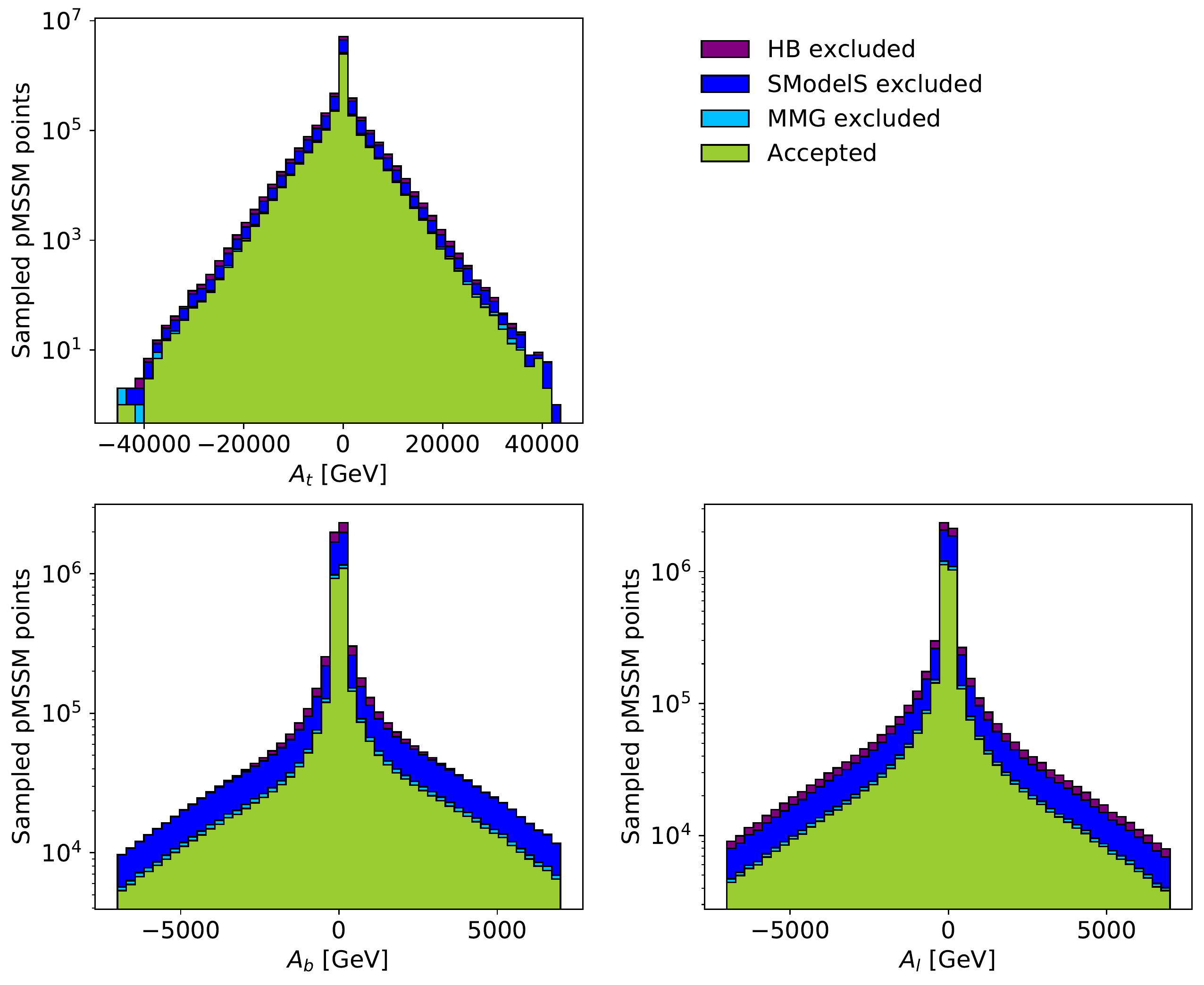}
   \caption{Distribution of the trilinear couplings $A_t$, $A_b$, and $A_l$ of sampled pMSSM points.}
  \label{pmssm-1}
\end{figure}

The distribution of the sampled trilinear couplings ($A_t$, $A_b$, and $A_l$) are shown in Figure \ref{pmssm-1}. The symmetry across 0 shows that the random selection of initial scan points does indeed populate all sign combinations effectively. 

\begin{figure}[htbp]
    \centering
    \includegraphics[width=0.9
    \textwidth]{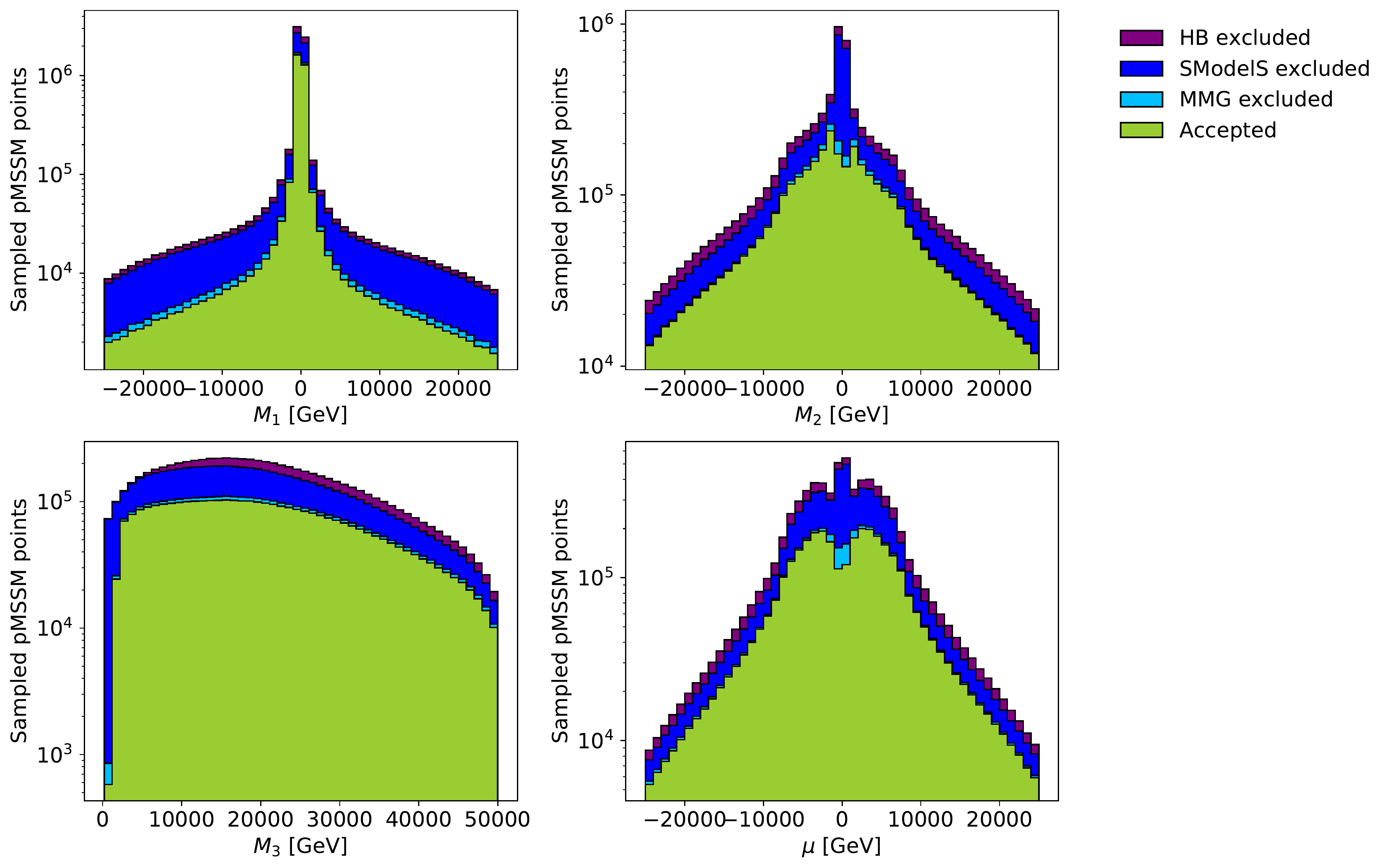}
    \caption{Distribution of the gaugino mass parameters $M_1$, $M_2$, $M_3$, and $\mu$ of sampled pMSSM points.}
    \label{pmssm-2}
\end{figure}

The distributions of the sampled gaugino mass parameters ($\mu$, $M_1$, $M_2$, and $M_3$) are shown in Figure \ref{pmssm-2}. Again, good symmetry across 0 is observed in $\mu$, $M_1$, and $M_2$. The power of LHC SUSY searches is visible at low values of these mass parameters, where the SModelS exclusion rules out the majority of points accepted by the McMC. 

\begin{figure}[htbp]
  \centering
  \includegraphics[width=0.7\textwidth]{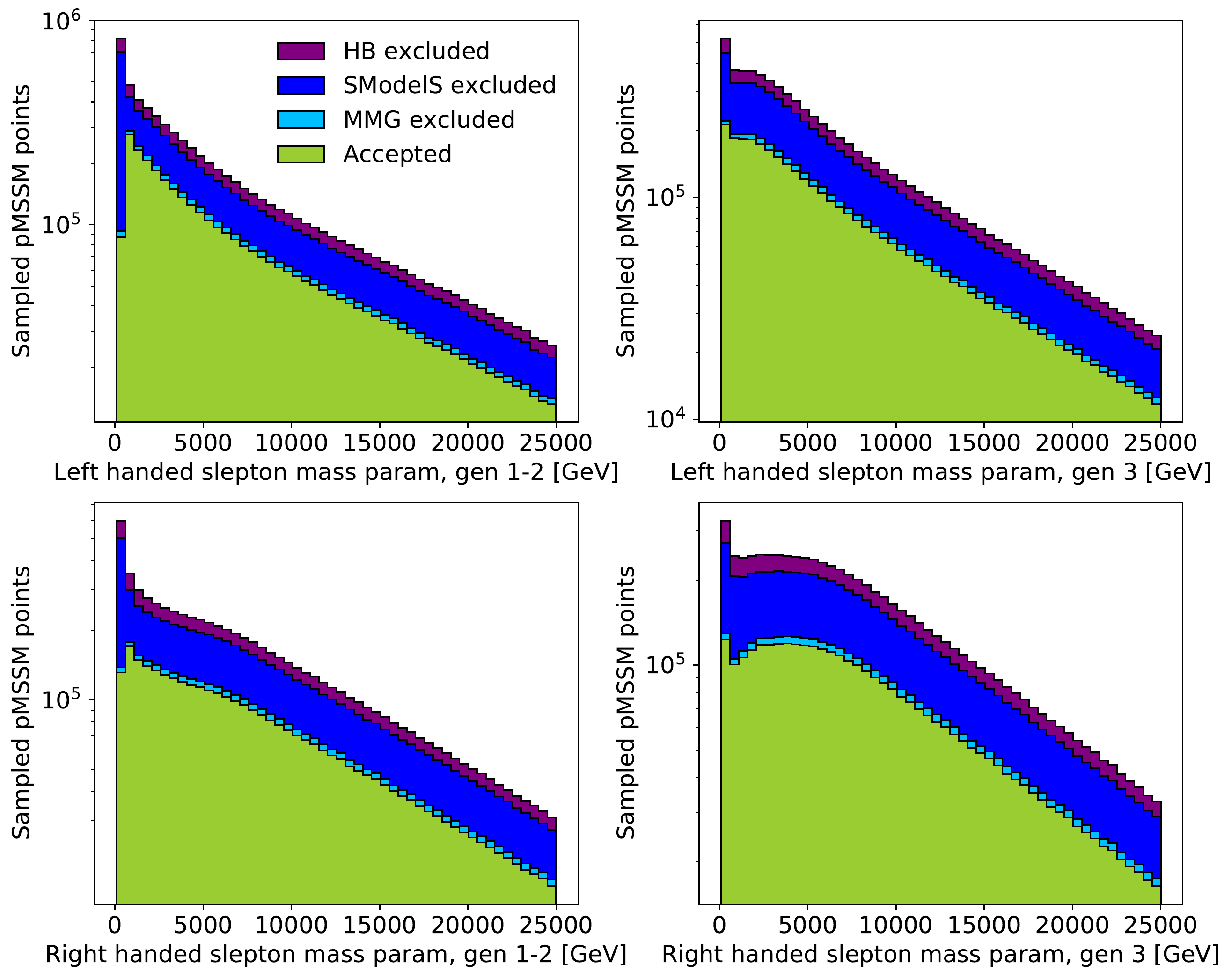}
  \caption{Distribution of slepton mass parameters $m_{\tilde{L}^{1,2}}$, $m_{\tilde{L}^3}$, $m_{\tilde{R}^{1,2}}$, and $m_{\tilde{R}^3}$ of sampled pMSSM points.}
  \label{pmssm-3}
\end{figure}

The sampled values of slepton mass parameters are shown in Figure \ref{pmssm-3} for left-handed (left) and right-handed (right) sleptons. The peak at low values of the smuon/selectron mass corresponds to those points with large $\Delta a_\mu$. These are largely excluded by SModelS. 

\begin{figure}[htbp]
  \centering
  \includegraphics[width=0.9\textwidth]{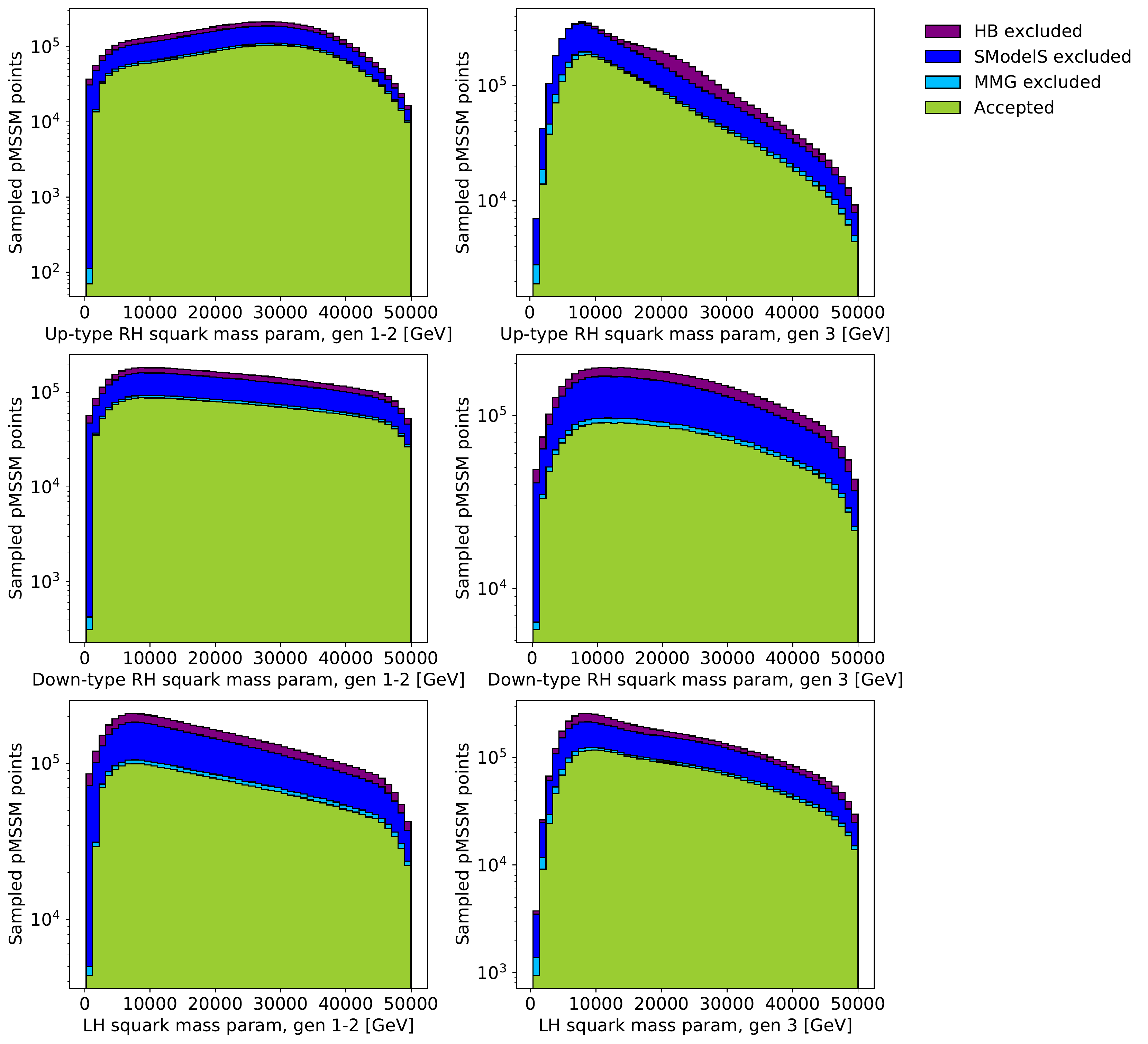}
  \caption{Distribution of squark mass parameters (top)  $m_{\tilde{u}^{1,2}}$, $m_{\tilde{u}^{3}}$,  (middle) $m_{\tilde{d}^{1,2}}$, $m_{\tilde{d}^{3}}$, (bottom) $m_{\tilde{q}^{1,2}}$, $m_{\tilde{q}^{3}}$ of sampled pMSSM points.}
  \label{pmssm-4}
\end{figure}

The sampled squark mass parameters are shown in Figure \ref{pmssm-4}: the left column shows the right-handed up-type, the center column shows the right-handed down-type, and the right column shows the left-handed squark masses. Again, LHC searches have been able to exclude much of the parameter space at low squark masses, as showwn by the SModelS exclusion. 

\begin{figure}[htbp]
  \centering
  \includegraphics[width=0.9\textwidth]{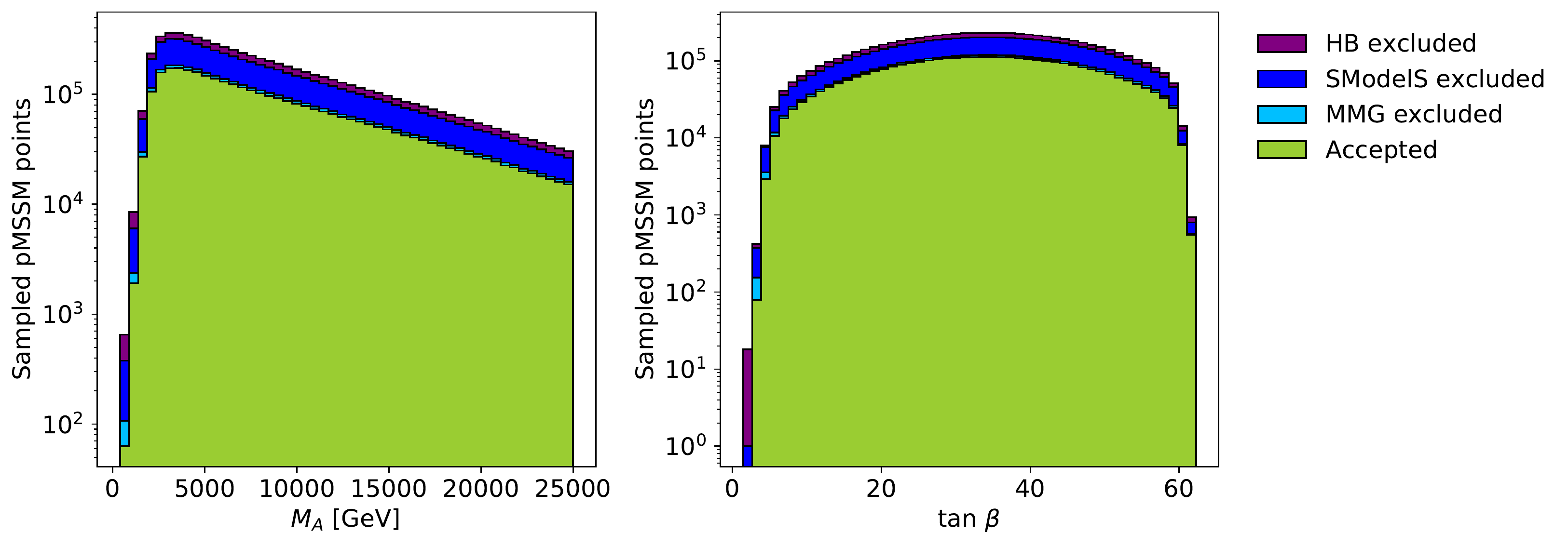}
   \caption{The sampled mass of the heavy Higgs boson $M_A$ (left), and $\tan\beta$ (right) for the sampled pMSSM points.}
  \label{pmssm-5}
\end{figure}

The pMSSM Higgs paramters, the mass of the heavy Higgs boson $M_A$ and $\tan\beta$, are shown in Figure \ref{pmssm-5}.  HiggsBounds, which includes dedicated LHC searches for the decay of the heavy Higgs, and SModelS exclude all points with $M_A<500$ GeV.

\FloatBarrier
\subsection{Sparticle masses}

\begin{figure}[htbp]
  \centering
  \includegraphics[width=0.7\textwidth]{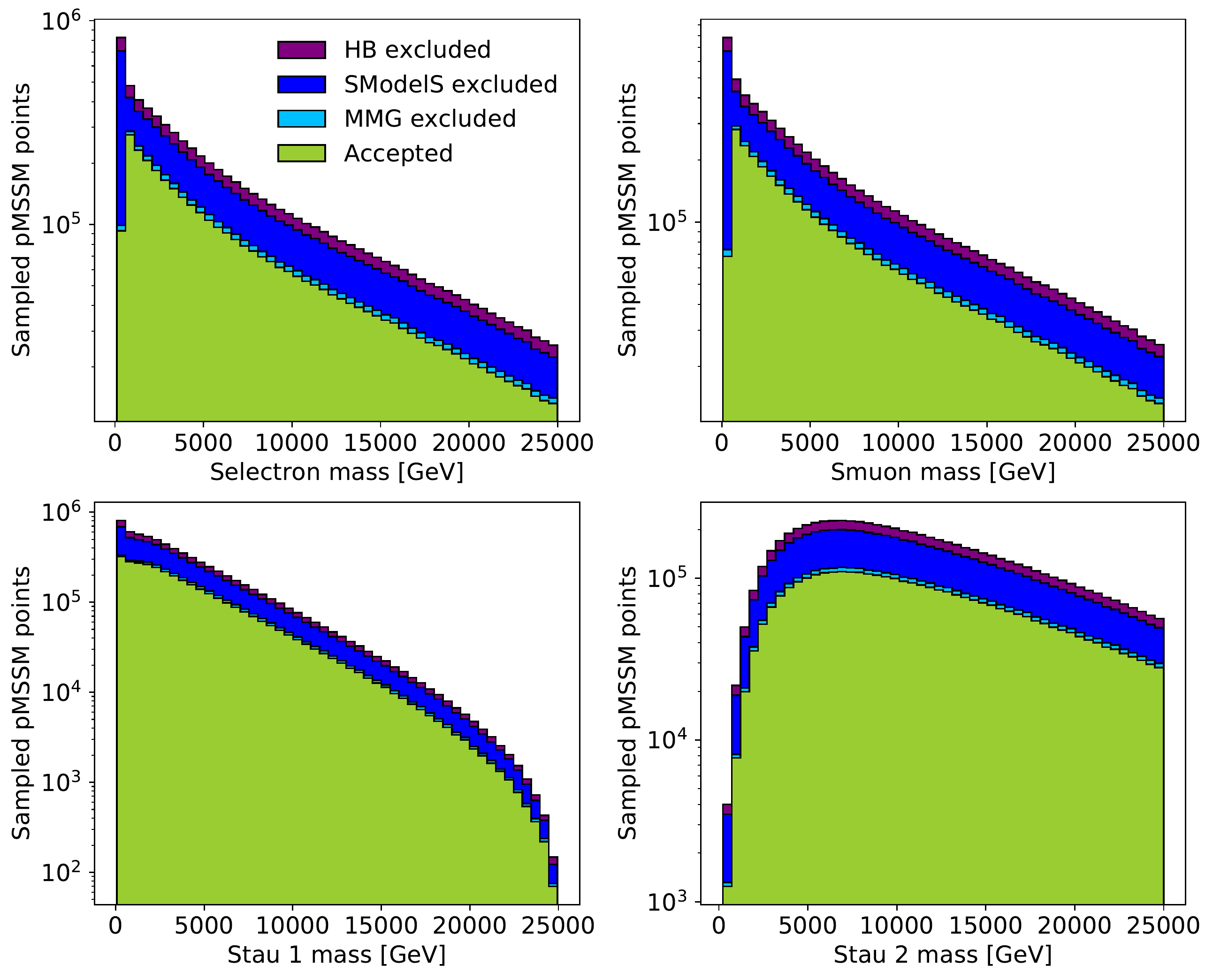}
   \caption{Slepton masses}
  \label{slepton-masses}
\end{figure}

\begin{figure}[htbp]
  \centering
  \includegraphics[width=0.95\textwidth]{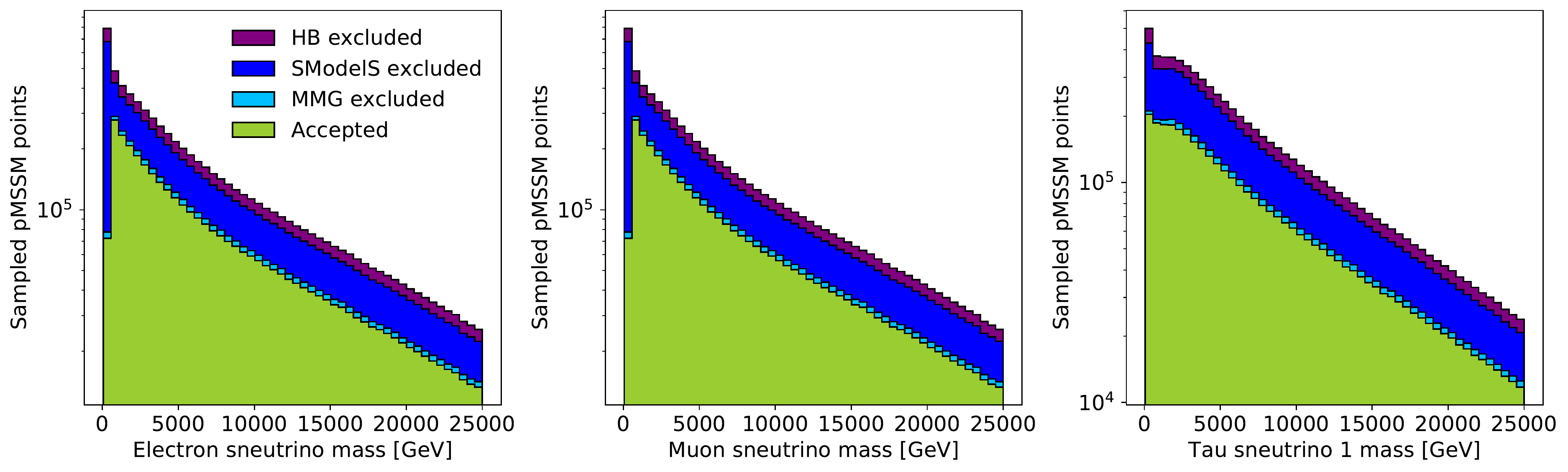}
   \caption{Sneutrino masses}
  \label{sneutrino-masses}
\end{figure}

\begin{figure}[htbp]
  \centering
  \includegraphics[width=0.9\textwidth]{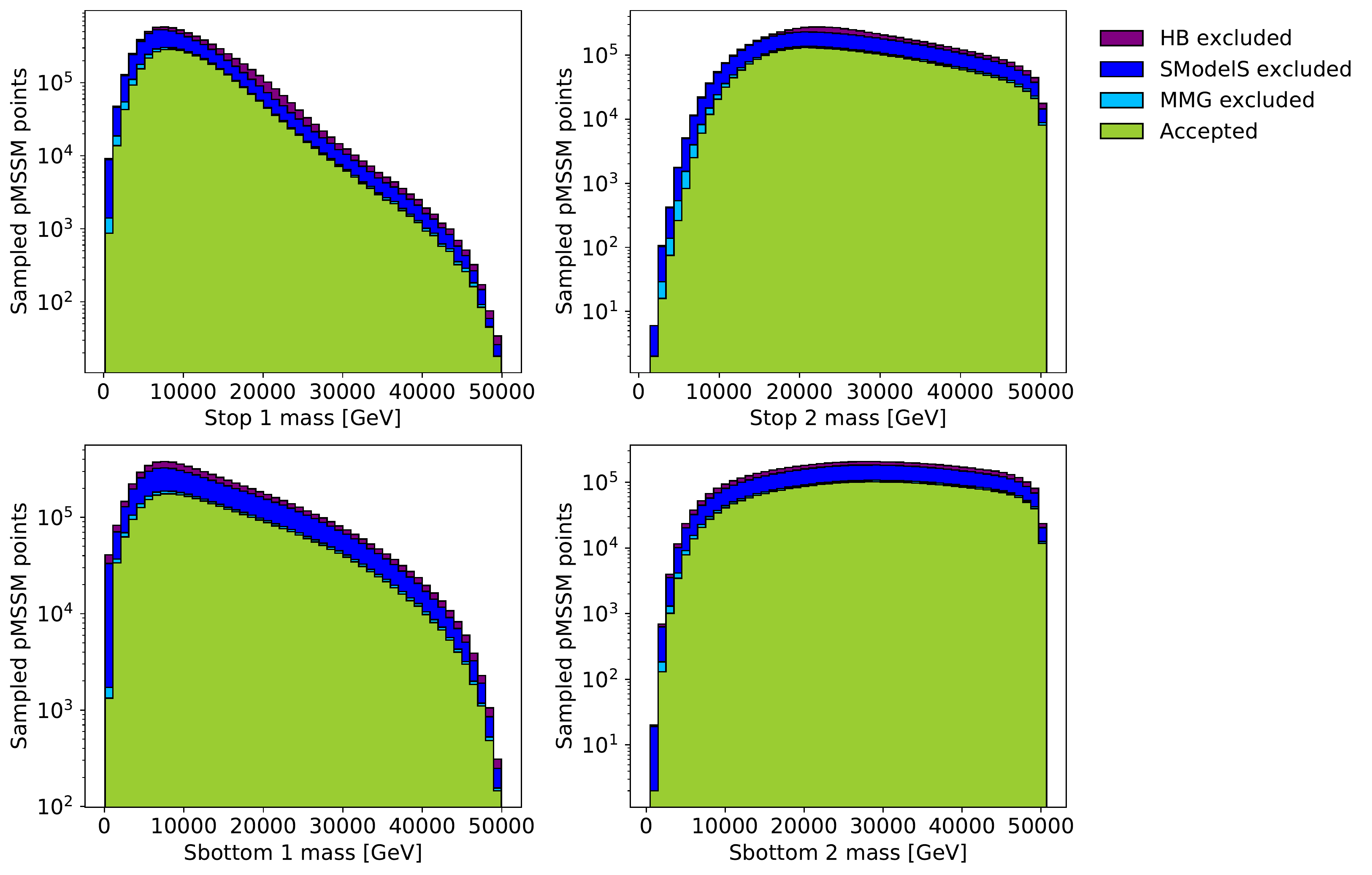}
   \caption{Heavy squark masses}
  \label{squark-masses}
\end{figure}

\begin{figure}[htbp]
  \centering
\includegraphics[width=0.7\textwidth]{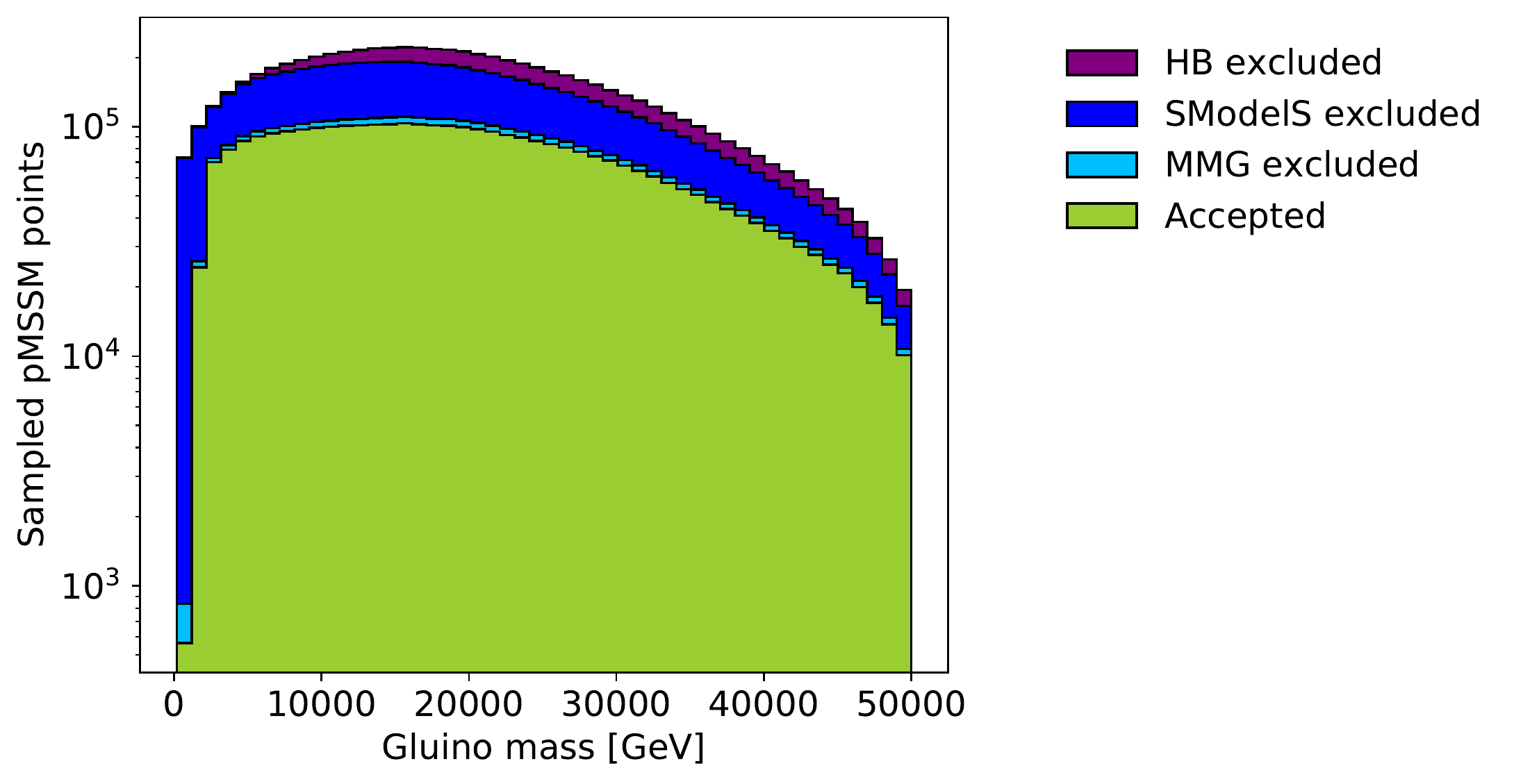}
   \caption{Gluino mass}
  \label{gluino-mass}
\end{figure}

\begin{figure}[htbp]
  \centering
  \includegraphics[width=0.9\textwidth]{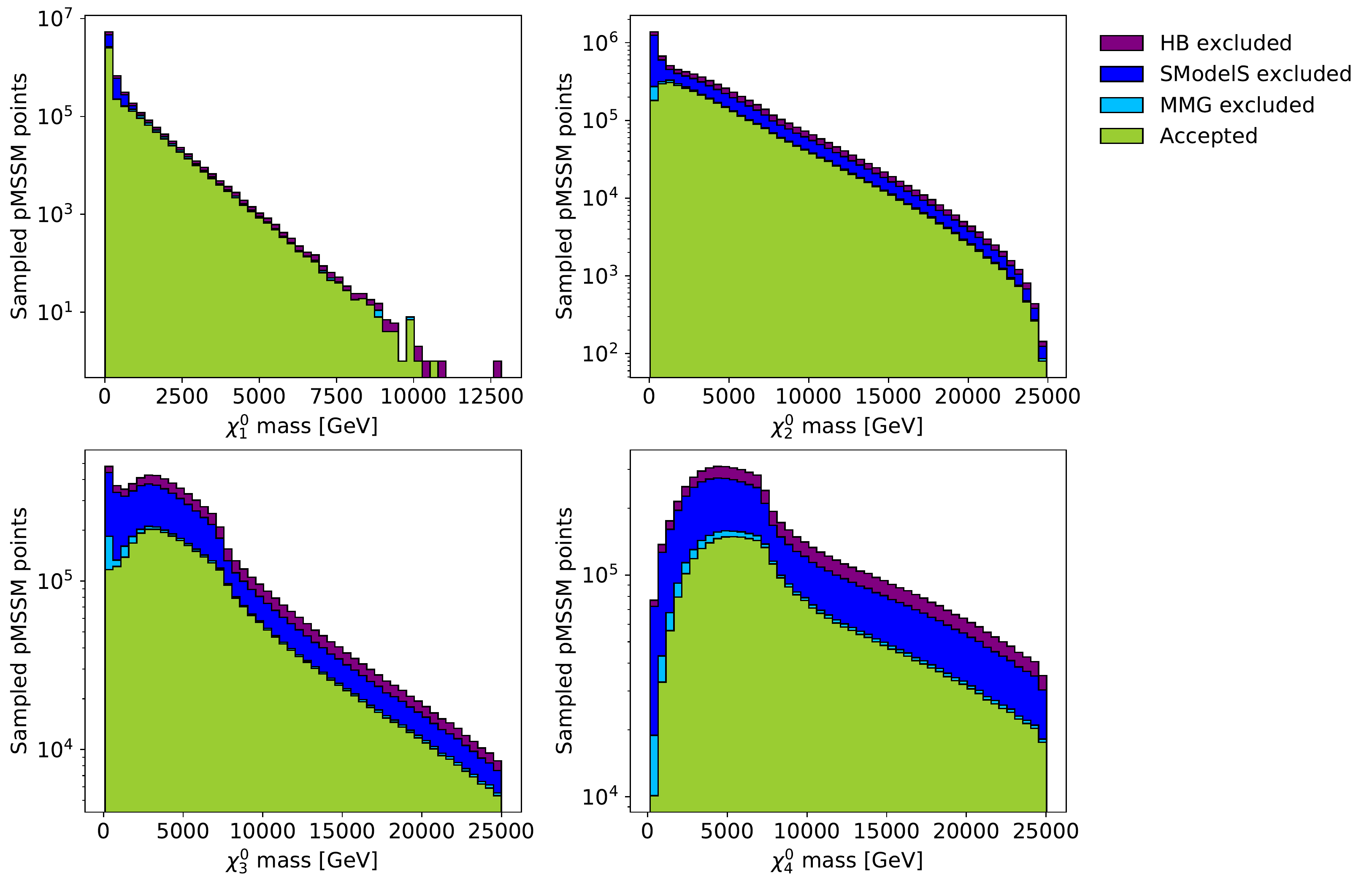}
   \caption{Neutralino masses}
  \label{neutralino-masses}
\end{figure}

\begin{figure}[htbp]
  \centering
  \includegraphics[width=0.9\textwidth]{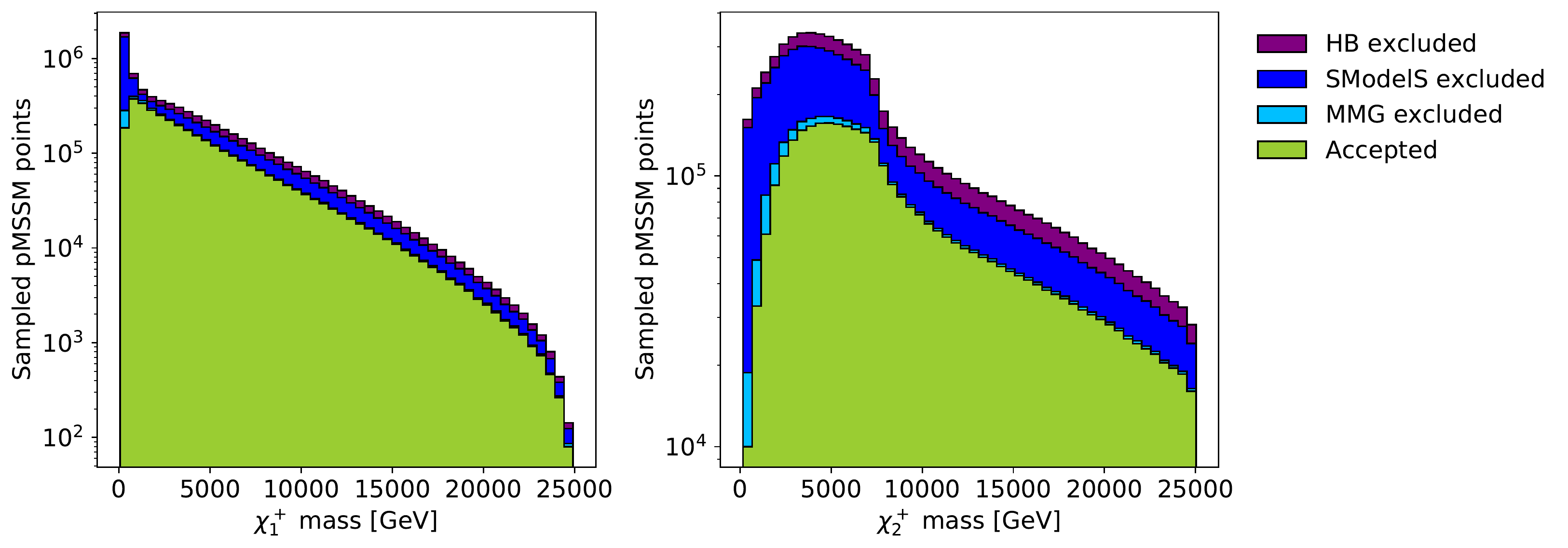}
   \caption{Chargino masses}
  \label{chargino-masses}
\end{figure}

This section shows the distributions of sparticle masses of the sampled pMSSM points.  Figure \ref{slepton-masses} shows the slepton masses, and Figure \ref{sneutrino-masses} shows the sneutrino masses. Figure \ref{squark-masses} shows the sbottom and stop quark masses, and Figure \ref{gluino-mass} shows the gluino mass.  The neutralino and chargino masses are shown in Figures \ref{neutralino-masses} and \ref{chargino-masses}, respectively.

\FloatBarrier
\subsection{Standard model observables}
\begin{figure}[htbp]
  \centering
  \includegraphics[width=0.7\textwidth]{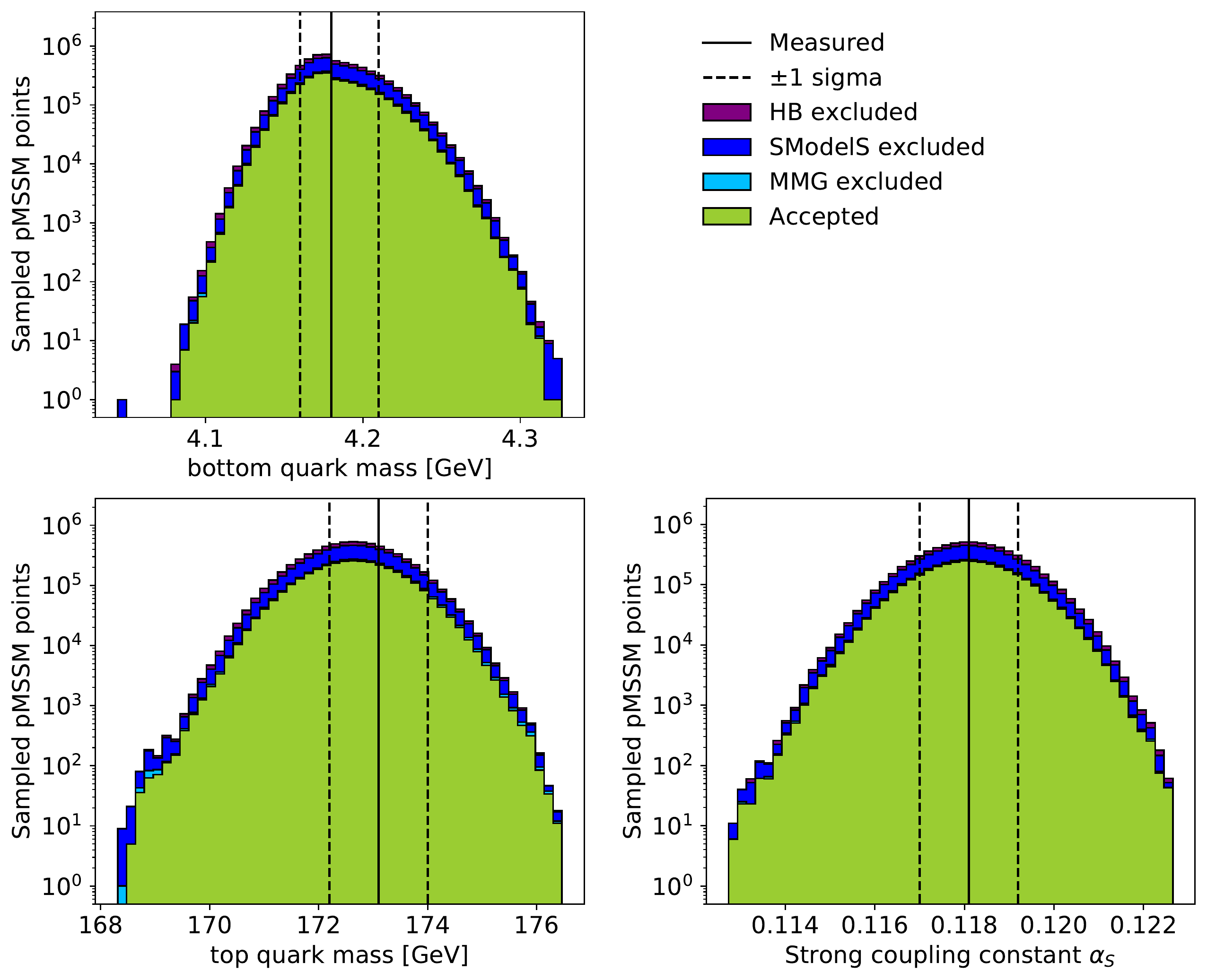}
   \caption{The bottom quark mass, top quark mass, and strong coupling constant for each sampled pMSSM point, with the measured value and uncertainties indicated by vertical lines.}
  \label{mtmbas}
\end{figure}

The top and bottom quark masses, along with the strong coupling constant $\alpha_S$,  for each pMSSM point are shown in Figure \ref{mtmbas}.  

%\FloatBarrier
%\subsection{Electroweakino dark matter}
%The composition of the dark matter particle in the sampled pMSSM points is shown in Figure \ref{dmcomposition}. The top left corner of each panel corresponds to pure wino, the bottom left to pure higgsino, and the bottom right to pure bino dark matter.  Most pMSSM points contain an LSP that is relatively pure in the electroweakino composition, though the scan also captures points with a mixture. 

%\begin{figure}[htbp]
%  \centering
%  \includegraphics[width=0.95\textwidth]{sections/scans/plots/dm_composition.pdf}
%   \caption{The electroweakino composition of the dark matter candidate in terms for each pMSSM point, (left) for points accepted by the McMC and (right) for points passing all post-processing selection. In each panel, the top left corner corresponds to pure wino, the bottom left to pure higgsino, and the bottom right to pure bino dark matter. }
%  \label{dmcomposition}
%\end{figure}
%\FloatBarrier

\end{document}